\documentclass[a4,11pt]{article}
\usepackage{epsfig}
\usepackage{amssymb,epsfig,graphicx,epsf,epstopdf}

\usepackage[titletoc,toc,title]{appendix}

\usepackage{amsfonts}
\usepackage{amsmath}
\usepackage{feynmf}

\usepackage{latexsym}
\usepackage{float} %Posiciona la imagen donde la pones el codigo exactamente.

\usepackage{slashed}
\usepackage{cite}
\usepackage[utf8]{inputenc} % acentos! Tildes en realidad
\usepackage{mathtools}
\usepackage[font=small]{caption}
\usepackage{color}
\DeclareFontFamily{OT1}{pzc}{}
\DeclareFontShape{OT1}{pzc}{m}{it}{<-> s * [1.10] pzcmi7t}{}
\DeclareMathAlphabet{\mathpzc}{OT1}{pzc}{m}{it}

\usepackage{commath} % Para módulo como \abs{}
\usepackage{accents} % Punto para la derivada \dt{} o \ddt{}
\usepackage{relsize} %agranda texto adentro de una ec. {\mathlarger{(x+y=z)}}
\usepackage[makeroom]{cancel} % Para tachar algo en una ecuacion
\usepackage{xcolor}
\usepackage[normalem]{ulem} %Permite subrayar varios renglones

\usepackage[colorlinks=true, allcolors=black]{hyperref} % Click in reference and goes there.

\usepackage{stackengine}
\usepackage{scalerel}

\usepackage{physics} %For bra ket notation

\newcommand{\ben}{\begin{equation}}
\newcommand{\een}{\end{equation}}
\newcommand{\be}{\begin{equation*}} 
\newcommand{\ee}{\end{equation*}}

\newcommand{\ba}{\begin{eqnarray}}
\newcommand{\ea}{\end{eqnarray}}
\newcommand{\bal}{\begin{aligned}}
\newcommand{\eal}{\end{aligned}}

\begin{document}
\title{Cosmological implications of the hydrodynamical phase of group field theory}

\author{
	Luciano Gabbanelli$^{\,a,}$\footnote{luciano.gabbanelli@uab.cat}$\ $ and Silvia De Bianchi$^{\,b,c,}$\footnote{silvia.debianchi@unimi.it\vspace{10pt}}\\[5ex]
$^a\,$\small Institut de Física d'Altes Energies (IFAE), Universitat Autònoma de Barcelona,\\ 
\small Campus UAB 08193 Bellaterra (Barcelona), Spain.\\[1ex]
$^b\,$\small Department of Philosophy, University of Milan, Via Festa del Perdono, 7\\
\small 20122 Milan, Italy.\\[1ex]
$^c\,$\small Department of Philosophy, Universitat Autònoma de Barcelona,\\
\small Building B Campus UAB 08193 Bellaterra (Barcelona), Spain.
\vspace{0pt}}

\date{}

\maketitle

\begin{abstract}
In this review we focus on the main cosmological implications of the Group Field Theory approach, according to which an effective continuum macroscopic dynamics can be extracted from the underlying formalism for quantum gravity. Within this picture what counts is the collective behaviour of a large number of quanta of geometry. The resulting state is a condensate-like structure made of “pre-geometric” excitations of the Group Field Theory field over a no-space vacuum. Starting from the kinematics and dynamics, we offer an overview of the way in which Group Field Theory condensate cosmology treats solutions for the homogeneous and isotropic universe. These solutions including a bounce, share with other quantum cosmological approaches the resolution of the singularity characterizing general relativity. Contrary to what is usually done in quantum cosmology, in GFT cosmology no preliminary symmetry reduction is needed for this purpose. We conclude with a discussion of the limits and future perspectives of the Group Field Theory approach.

\vspace{15pt}

\end{abstract}

\section{Introduction}

The goal of the present paper is to review some essential properties of Group Field Theory (GFT) and GFT condensate cosmology in an accessible way from a non-specialist’s perspective, and with a non-GFT audience in mind. For a review with an internalist perspective see \cite{ReviewCosmology}. Our purpose is to underline the physical consequences of the formalism and to discuss some results that go beyond quantum cosmology and invite conceptual innovation.
Although the general formalism of GFT is deeply related to Loop Quantum Gravity (LQG), tensor models and lattice quantum gravity, in the sense that all these theories share the formulation in terms of discrete and combinatorial structures rather than continuum variables (such as a metric or a connection on a smooth manifold), the peculiarities introduced by the cosmological sector of GFT brings a novel proposal on how to deal with these fundamental structures in order to recover a continuum spacetime and ultimately general relativity \cite{Oriti2014}.

Furthermore, GFT condensate cosmology is able to reach results similar to those of Loop Quantum Cosmology (LQC) without any symmetry reduction, namely generic solutions of its models of the early universe include a bounce.
Symmetry reduced versions of LQG, well known as LQC, have recently been studied as models of quantum cosmology \cite{LQCNoSingularity}. 
They possess two main features. The first is a mechanism for avoiding the big bang singularity in the framework of mini-superspace models of quantum gravity\footnote{
It should be noted that ‘singularity avoidance mechanisms’ may exist in more conventional mini-superspace of quantum geometrodynamics. For instance, from simple particle models like \cite{Russo} to more comprehensive studies in more realistic situations \cite{Bouhmadi-Lopez.Kiefer.Sandhoefer.Vargas2009} and recent extension to anisotropic models \cite{Kiefer.Kwidzinski.Piontek2019}, in which the analysis is consistently based on the behaviour of the wavefunction and not on the bouncing behaviour of quantum-corrected classical equations. For an overview and a comparison between LQC and standard quantum cosmology, see \cite{Bojowald.Kiefer.Moniz2010}.}. 
In this mechanism the inverse scale factor is represented by an operator that stays bounded as the universe's classical radius shrinks to zero. Other alternatives involve the effective discretization of the Hamiltonian constraint, which enables the quantum wavefunction to ‘jump over’ the singularity. Whichever the model under consideration, it is not clear how these models can be derived from full-fledge LQG, if it is in fact possible. Hence, there is no common agreement whether the singularity avoidance is a property of the full theory. In fact, calculations regarding the full LQG theory show that the spectrum of the operator corresponding to the inverse volume is not bounded from above \cite{LQGNoBound}\footnote{For further analysis concerning these topics refer to \cite{ContrastingQuantumCosmologies}.}.

The second feature implies the possibility that an intrinsically quantum gravitational mechanism of LQC might trigger inflation, which may eventually be stopped (gracefully) by gravity itself. Although proposals have been made in the context of isotropic models \cite{InflationQG}, in LQC there is no clear success in producing an inflationary era out of pure quantum gravity effects. On the contrary, in GFT cosmology seems to be possible by modeling an early epoch of accelerated expansion, which can last for an arbitrarily large number of e-folds without the need of introducing an {\it ad hoc} potential for the scalar field \cite{GFTInteracting}.

The same key-feature of LQC including a bounce can be found in GFT cosmology. 
According to GFT, cosmology is understood as the `hydrodynamic' regime of quantum gravity. The macroscopic universe would correspond to a fluid whose `atoms' are GFT quanta, behaving collectively in a mean field approximation. The collective variable is a sort of density function, to which a velocity function is added \cite{UniverseAsQGC}.
Therefore, this `hydrodynamic' approximation provides a natural link with the usual Wheeler–DeWitt approach to quantum cosmology, but with two main improvements: (i) the theory does not require any symmetry reduction, but requires a suitable coarse-graining from the microscopic features; and (ii) includes many-body features of the full Hilbert space of the microscopic theory to the extent in which they survive the coarse-graining procedure and have an actual signature in the collective description. Nonetheless, there is no much control over whether the approximation survives the inclusion of quantum fluctuations of observables and inhomogeneous perturbations. In fact, \cite{MarchettiOritiQuantFluc} claims that the hydrodynamic approximation itself can break down in the regime of observables corresponding to the bounce where quantum fluctuations seem to become larger. In this scenario, the fact that expectation values of the volume and the density show a bouncing dynamics becomes less relevant, calling for a more refined approximation of the underlying quantum gravity dynamics.

The idea of modeling the universe as a GFT condensate is part of a larger effort aimed at understanding the collective dynamics of many interacting degrees of freedom.
If one could find a physical significance of the non-spatiotemporal quanta of the theory, then one would wonder which are the possible macroscopic phases on which the quanta can be organized. Therefore, it seems plausible that geometry and cosmology as we know them emerge in (at least) one of these phases.
In other words, spacetime and the universe emerge through a phase transition from some non-geometric phase (or pre-geometric phase) with no notion of locality. This is consistent with the aim of background-independent formulation for quantum gravity theories. This phase transition has been dubbed {\it geometrogenesis}.
In the context of GFT, the condensate phase of the universe would arise from this {\it geometrogenesis} \cite{UniverseAsQGC,Oriti2007,Oriti2018}. This realization is supported by results on GFT renormalization suggesting this type of phase transition (see Ref. \cite{OritiRenormalization} for recent computations).
The step from the background independent and non-spatiotemporal microscopic quantum description to the effective cosmological one, involves a coarse graining of an infinite dimensional set of quantum interacting degrees of freedom. This situation is analogous to the effective hydrodynamic phase of some condensed matter system, which is obtained directly from the quantum field theory describing the atoms that constitute it.
Within this condensate phase the universe would ultimately enter in its hydrodynamical regime, where a more regular and ordered phase is achieved.
At this stage, the theory can be formulated in the language of spacetime, and geometry can be identified together with notions of locality, which are recovered by using relational observables.

We will focus on the GFT formalism according to which the macroscopic, homogeneous and isotropic universe dynamically emerges from the collective behaviour of a highly coherent configuration of many discrete ``pre-geometric atoms" introduced in Sections \ref{QGmatterRF}. 
This approach suggests the existence of a phase where a large number of quanta condensate, and ultimately within this condensate phase, the system enters in a `hydrodynamic regime' where concepts of space and time are well defined (see Section \ref{CondensatePhGFT}). In this regime, the classical Friedmann dynamics for a homogeneous and isotropic universe, together with quantum corrections of general relativity, emerge consistently from the fundamental constituents. These aspects together with deviations from exact homogeneity and the construction of more realistic cosmological scenarios are presented and discussed in Sections \ref{FrdmnUniv} and \ref{Rods} of our review. We then conclude by discussing the results in recent GFT literature and their cosmological implications, underlying open questions to be further investigated.

\section{Quantum gravity with matter reference frames}\label{QGmatterRF}
GFT is a research programme for a non-perturbative quantization of gravity. These theories aim at describing the dynamics of quanta of space on background-independent theories and hence are characterized by a lack of any preferred notion of time. 
According to GFT, the universe is an ensemble of processes happening where any notion of evolution is purely relational. 
Hence, as it happens in most quantum gravity approaches, discussions focus on how to identify in mathematical terms the available degrees of freedom at the Planck scale in order to define the relational dynamics.

One of the most drastic change of perspective of GFT is certainly how the theory describes the macroscopic universe starting from the underlying physics and without referring to any external structure. 
According to this approach, the structure and dynamics of “quanta of space”, identified with discrete “pre-geometric” elementary structures (usually, quantized tetrahedra when restricted to 4 dimensions) each of which with an associated classical phase space, provide the `appearance' of the spacetime fabric. 
Indeed, one might consider the classical theory as an emergent phenomenon that agrees with general relativity, together with the diffeomorphism invariance which is one of its most established foundations.

As we will discuss below, this view is similar to the theory of superfluidity where the fundamental quantum atoms play no individual role at the hydrodynamic level, but the collective behaviour is what matters. Analogously, in the GFT formalism the macroscopic universe emerges dynamically as the collective behaviour of a highly coherent configuration of many discrete ``pre-geometric atoms". Thus, the theory enters into a ``hydrodynamical" phase where a large number of constituents form a condensate structure. It is from the latter that concepts of space and time are defined. This approach would suggest that, in this regime, the classical Friedmann dynamics, together with quantum corrections emerge consistently from the fundamental constituents. 

Let us first focus on gravity only with no mater fields present at all and later, we will show how this approach can embody matter degrees of freedom. 
In the formulation of GFT, the elementary degrees of freedom of geometry are represented by excitations of a quantum (statistical) complex scalar field $\varphi(g_1,g_2,...,g_d)$, function of $d$ group elements of an abstract group manifold (or the corresponding Lie algebra); this is
\begin{equation}\label{GFTFieldINd}
	\varphi(g_I):\hspace{10pt}G^d\quad\longrightarrow\quad\mathbb{C}\end{equation}
for $I=1,...,d$. As mentioned, this manifold does not carry {\it a priori} any notions of spacetime geometry by itself, but stores geometric information --metric or connection data-- beyond its mere combinatorial or topological structure \cite{GFT}. Then, the elementary excitations occur above a fully degenerate ‘no-space’ vacuum and can be seen as quanta of geometry labelled by data in the domain space of the bosonic GFT field $\varphi$.
Each quantum can be represented graphically as a $(d-1)-$simplex with field arguments associated to the faces of it, or as a $d-$valent graph vertex, with field arguments associated to the links. The dynamics is governed by the choice of the action $S(\varphi)$ that will be defined in Eq. \eqref{Action}. By appropriate choices of the dimension $d$, the group manifold $G$ and the functional form of the action together with the combinatorial pairing of field arguments, these theories can be understood as quantum field theories of spacetime \cite{GFTSFM}.

Concretely, in the cosmological context, most 4 dimensional gravity models use the spacetime dimension $d=4$ and the group $G=SU(2)$.
Indeed, when the GFT field 
\begin{equation}\label{GFTField}
	\varphi(g_I):\hspace{10pt}\text{SU}(2)^4\quad\longrightarrow\quad\mathbb{C}\end{equation}
satisfies the ``closure" condition $\varphi(g_I)=\varphi(hg_I)$ for each $h\in G=$ SU(2), the microscopic theory can be depicted as $3-$simplices, i.e. tetrahedra, whose 4 faces are associated to the field arguments given by an equivalent class of geometrical data $[\{g_I\}] = \{\{hg_I\}, h\in G\}$. By these appropriate choices, the perturbative expansion of the theory produces amplitudes that can be seen as a simplicial gravity path-integral \cite{BaratinOriti2012}, with the group-theoretic data entering as holonomies of a discrete gravitational connection.

Analogous algebraic data is used to construct the spin network states in LQG (holonomies of a connection and fluxes of a triad field) \cite{LQG}, and in fact these states can be seen as graphs dual to the triangulations formed by GFT quanta. In this duality, each vertex of the graph is dual to the tetrahedron of the triangulation; the links joining vertex are coloured by SU(2) connections and play the role of the tetrahedron faces where the gluing determines the particular GFT model. Therefore, GFT quantum states are built up from the kinematical data of LQG and the theory can be understood as a field-theoretic $2^{\text{nd}}$ quantization formulation of LQG \cite{2ndQuanOfLQG}. This correspondence between discrete quantum field theory (QFT) structures and spin networks can be found also at the dynamical level but treated via standard QFT methods. In this way, GFT attempts to define a sum over discretised geometries which can be used, once a continuum limit is identified, to obtain a path integral formulation for quantum gravity.

However, to extract this effective continuum physics and realistic cosmological models requires more crucial ingredients. Of course, there is plenty of matter in the universe and the relation of the matter content and its corresponding interaction must be addressed in a theory of Quantum Gravity. In GFT and other related diffeomorphism-invariant formalisms, matter fields are the most convenient way to define physical reference frames. This is a relational approach usually employed to define physical observables in different quantum gravity theories (see \cite{RelationalObservables1,RelationalObservables2,DustFrame}).

A standard choice in quantum cosmology is to use free massless scalar fields for defining the evolution of the theory. This choice ensures diffeomorphism invariance. It is our interest here to see how these scalar fields can be coupled to a QFT formalism and their implications for the GFT condensate and the cosmological sector.
The matter reference frame should be reconstructed from the physical degrees of freedom of the underlying theory. Therefore, the GFT field $\varphi$ in \eqref{GFTField} encompasses the new ``coordinate" (scalar) degrees of freedom with real labels,
\begin{equation}
	\varphi(g_I,\phi^J):\hspace{10pt}\text{SU}(2)^4\times\mathbb{R}^4\quad\longrightarrow\quad\mathbb{C} \,.\end{equation}
As before, the group elements $g_I\in$ SU(2) can be associated to the parallel transport of a gravitational Ashtekar–Barbero connection across the four faces of the tetrahedron, or equivalently along the links attached to each node of the $4-$valent spin network and dual to such faces.
Each ``chunk of space" is labelled with a $\phi^J$, with $J=0,1,2,3$, specifying the discrete matter (scalar field) degrees of freedom\footnote{This procedure is usually generalized to an arbitrary $k-$number of massless scalar fields, $J=0,1,..., k$. Here we restrict the analysis to only four of them because they will be used for labelling the four spatiotemporal dimensions; i.e. 1 temporal $\phi^0$ and 3 spatial $\phi^i$ independent components.}. These fields are attached to the vertices corresponding to each tetrahedron and would represent the readings of all fields: `clocks' and `rods'.

The $2^{\text{nd.}}$ quantization formalism is suitable for describing quantum many-body systems. The Fock space is built from the Fock vacuum $\ket{0}$ representing the state with no spin network nodes or no tetrahedra. Therefore, it is a state with no topological nor geometrical information; a ``no-space" vacuum analogous to the Ashtekar–Lewandowski vacuum \cite{Ashtekar.Lewandowski1993} where operators for geometric observables such as volumes and areas from LQG vanish.
Thereupon, one-particle states can be generated with the creation operators $\hat\varphi^\dagger(g_I,\phi^J)$ acting on the vacuum state. 
Both ladder operators, $\hat\varphi^\dagger$ together with the annihilation operator $\hat{\varphi}$, are required to be translation invariant under diagonal group multiplication from the left
	\begin{equation}\label{GaugeLeftInvariance}
		\varphi(g_1,\dots,g_4,\phi^0,\dots,\phi^3)=\varphi(hg_1,\dots,hg_4,\phi^0,\dots,\phi^3) \qquad\forall h\in \text{SU}(2)\,.\end{equation}
It can be verified that if the theory is written as a field theory on the Lie algebra $\mathfrak{su}(2)^4\simeq(\mathbb{R}^3)^4$ via a noncommutative Fourier transform, the counterpart of the left gauge invariance is the closure constraint for the four faces of the tetrahedron \cite{BaratinDittrichOritiTambornino}. The role of the ladder operators is derived directly from the postulated canonical commutation relations for the chosen bosonic statistics, which include the correct left invariance. These are
\begin{equation}
	\Bigl[\hat\varphi(g_I,\phi^J),\hat\varphi^\dagger(g'_I,\phi'^J)\Bigr]=\delta^4(\phi^J-\phi'^J)\int_{\text{SU(2)}}\text{d}h\,\prod_{I=1}^{4}\delta(g'_I\,h\,g_I{}^{-1})\ ,\end{equation}
while two $\hat\varphi$ or two $\hat\varphi^\dagger$ operators commute. 
This leads us to interpret $\hat\varphi^\dagger$ as a creator of a single four-valent spin network node/tetrahedron with data given by $g_I$ up to a gauge transformation on the left.
In this picture, the one-particle state is depicted as

\vspace{2pt}\begin{minipage}{0.7\textwidth}\begin{equation*}
\hat\varphi^\dagger(g_I,\phi^J)\ket{0}\hspace{3pt}=\hspace{3pt}\ket{g_I,\phi^J}\hspace{3pt}=\end{equation*}\end{minipage}\hfill\begin{minipage}{0.8\textwidth}\vspace{-10pt}\begin{figure}[H]\hspace{-80pt}\includegraphics[scale=0.1]{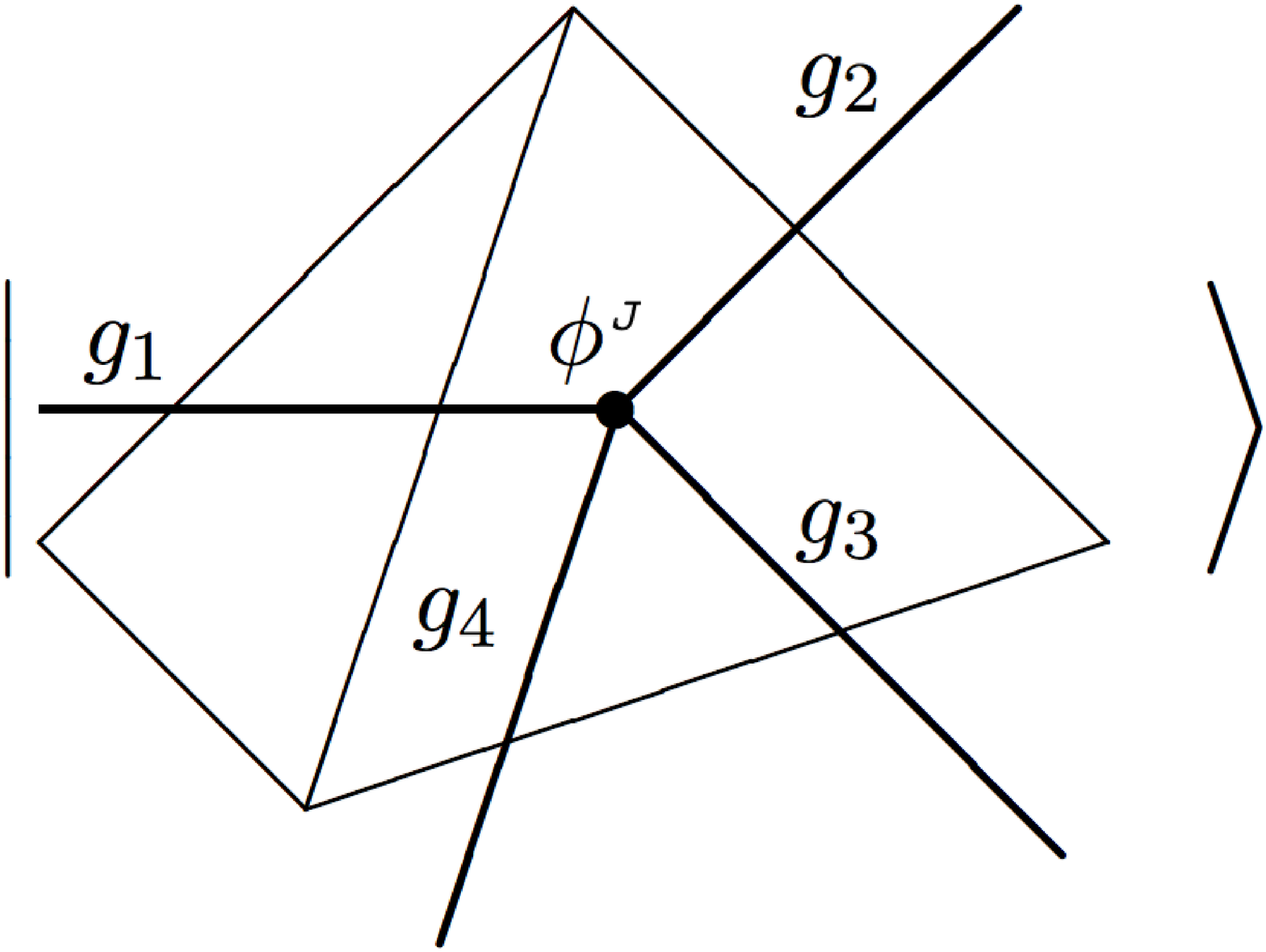}\end{figure}\end{minipage}\vspace{5pt}
Finally, one can create generic multiparticle states with arbitrary particle number $N$ acting $N$ times with the creation operator $\hat\varphi^\dagger$ on $\ket{0}$.

Generic $N-$particle states can be associated to graphs structures and therefore, GFT models can be understood as a QFT for spin networks. 
Even though these states are analogous to those postulated in LQG in the sense that the degrees of freedom are encoded in purely combinatorial and algebraic structure as functions on these group manifolds, the corresponding Hilbert spaces are very different from each other.
The GFT Hilbert space has a Fock structure defined out of the union of graph-based Hilbert spaces decomposed into elementary building blocks.
An important feature of these Fock states is that, contrary to LQG states which contain all possible graphs with precise equivalence relations among them, they are defined with no unambiguous identification with the graph used to be constructed. They are superpositions of states for different particle numbers, and hence rather different from the states usually considered in LQG which are typically built on fixed graphs.
Furthermore, although both theories define states with different numbers of nodes as orthogonal, the same does not happen when the number of nodes coincides. Contrary to LQG states, the structure of GFT states is reduced to specific correlations among fundamental quanta. In this manner, quantum states with the same number of quanta but with different correlations can overlap. Therefore, one could say that the structure of the GFT graphs has less relevance than in standard LQG.
The precise connection between the (kinematical) LQG states and GFT states is a delicate issue and we refer to \cite{Hilbert} for a more complete discussion. However, two more comments are worth mentioning here.
First, in 2$^{\text{nd.}}$ quantization formalism, the very number of particles corresponding to Fock states is uncertain, and becomes an observable with probabilities and mean values as any other observable. Therefore, as it will be discussed in the following in Eq. \eqref{NumberOp}, $N$ is a quantum observable of the theory and the dependence on it is derived and not assumed, and enters necessarily in both the kinematics and the effective cosmological dynamics \cite{QCfromQGCLattice}.
Moreover, although graph structures lose importance in GFT, their physical relevance is enhanced because the number of graph nodes is turned into a new (very simple) physical observable. 
If one takes seriously the features of GFT states with their fundamental discreteness and drops some of the features corresponding to the LQG Hilbert space, as stated in \cite{Hilbert}, ``the GFT Hilbert space has a clear Fock structure, giving straightforward meaning to the notion of ‘QG atom of quantum space’, and making powerful analytical tools available".

Second, if cosmology and geometry emerge from an effective hydrodynamic approximation to quantum gravity, it is plausible that topological information can not be found in the representation of Fock states in terms of graphs, but it must be extracted from elsewhere.
Stated in other words, the topological information that one may associate to a collection of $N$ tetrahedra plays no role in the continuum interpretation (and there would probably be no consistent way of using this information if we are dealing with superposition of states with different $N$), hence the topology of space should also be emergent rather than determined by microscopic details, such as specific choices of graphs. In fact, the continuum limit which, as briefly mentioned in the Introduction can be understood as a thermodynamic limit, is achieved through a phase transition reached in the $N\rightarrow\infty$ limit\footnote{The fact that $N\rightarrow\infty$ corresponds to a continuum limit in which a phase transition is reached has been shown in detail in matrix models for two-dimensional gravity \cite{FrancescoGinspargZinn}.}. Particularly noteworthy is that the resulting states after this phase transition are no longer in the original GFT Fock space. This is a standard feature in QFT: when phase transitions occur, a change of representation to a different, unitarily inequivalent Hilbert space is needed. Among the more general considerations that can be developed there is the following. If there is a theory with physical discreteness, one can argue that no continuum limit is realized, but only a “continuum approximation” in analogy with the one appearing in causal sets \cite{HensonCausalSet}. The GFT condensate that may describe the universe would then have an extremely large but still finite number of quantum gravity atoms which are approximated as infinite.

The GFT framework treats polyhedra quite literally as the quanta of spacetime. Quantum states in the kinematical GFT Hilbert space can be viewed as a collection of polyhedra (tetrahedra when $d=4$), which can be (or not) glued to each other across faces.
The theory is defined as a quantum field theory for the quanta where their gluing constraints together with their evolution processes can be interpreted as a definition of their interactions and their dynamics. Different choices for the interaction terms ${\cal V}$ are model-building choices and can be understood as different manners of defining generic non-local many-body interactions between the underlying quanta. Any such model building strategy should be based on a clear understanding of how simplicial geometry is encoded in the algebraic data that we used. Working with simplicial complex means to choose the dimension $d$ to the would-be spacetime dimension and interprets the GFT fields (this is, the quanta they create/annihilate) as $(d-1)-$simplices which its $(d-2)-$faces carry the arguments of the GFT fields, which for most practical purposes, one actually chooses the be the rotation subgroup \eqref{GFTField}. 
Although the theory admits a generic formulation in terms of simplicial polytopes whose boundaries are made of simplices, most of the relevant literature in models of quantum gravity and particularly the cosmological models relies on the well-understood simplicial case \cite{FriedmannEmergent,ReviewCosmology,UniverseAsQGC,ScalarField}. In the simplicial context, the simplicial gluing determines the face sharing interaction terms of five $3-$simplices (tethrahedra) to form a $4-$simplex, which are the only ones entering in the action; crucial for this interpretation is their non-locality in the group variables (in the sense that only a subset of the arguments of a given GFT field is related with a subset of the arguments of a different one). As mentioned, this kind of simplicial interactions is typically assumed in quantum geometric GFT models, but the same gluing process can be encoded in terms of dual graphs, understood as the $1-$skeleton of the cellular complex, dual to the simplicial complex of interest. Obviously, the Hilbert space taken by GFT also contains states associated to open spin network vertices not glued to any other, i.e. with some links (possibly all if interactions are neglected) ending up in $1-$valent vertices.
A simple general GFT action can be defined adding a quadratic kinetic term, that includes a local kinetic operator ${\cal K}$ containing derivatives with respect to both variables $g_I$ and $\phi^J$, to a generic interaction term  ${\cal V}$, which as mentioned is higher order in the field operators
\begin{equation}\label{Action}
	S[\varphi,\bar\varphi]=-\int_{\text{SU}(2)^4\times\mathbb{R}^4}\text{d}^4g\,\text{d}^4\phi\ \bar\varphi(g_I,\phi^J)\,{\cal K}\ \varphi(g_I,\phi^J)+{\cal V}[\varphi,\bar\varphi]\ .\end{equation}

Once the action is specified, the full dynamics is determined. In fact, the action is typically chosen specifically such that the perturbative expansion of the partition function of the field theory 
\begin{equation}
	{\cal Z}=\int{\cal D}\varphi{\cal D}\bar\varphi\, e^{-S[\varphi,\bar\varphi]}\end{equation} 
generates the Feynman rules for any spin foam model; such perturbative expansion in Feynman diagrams equals the sum over discretized path integrals for quantum gravity \cite{GFTSFMPerez}; in this sense is that GFT represents a $2^{\text{nd}}$ quantized reformulation of the LQG state space and a completion of the spin foam formalism.
The kernels ${\cal K}$ and ${\cal V}$ determine the details  of the resulting Feynman amplitudes and in doing so, it is possible to generate models that are related in a precise way to spin foam models \cite{Perez2013}.
Indeed, in \cite{Pietri.Freidel.Krasnov.Rovelli2000} it was realized that amplitudes for the Barrett–Crane spin foam model in four dimensional quantum gravity \cite{Barrett.Crane2000} could be obtained from a suitable choice of the GFT action. Later, it was shown that any prescription for a spin foam amplitude (within a class of models of interest for quantum gravity) could be obtained directly from GFT \cite{GFTSFM}.
In fact, the generality on possible choices of both operators, ${\cal K}$ and ${\cal V}$, points towards a one-to-one correspondence between spin foam models and GFT actions \cite{Pietri.Freidel.Krasnov.Rovelli2000}, where the GFT partition function ${\cal Z}$ corresponds to a sum over topologies and spacetime histories (for gravity and matter). Each history itself is discrete and contains a finite number of degrees of freedom. The main technical challenges are still the same as for the lower-dimensional matrix models; these are to control the unwieldy sum over Feynman graphs and obtain a continuum limit. However, we are interested here in cosmological implications of GFT, hence we will not go deeper in the microscopic description of the theory and proceed to the effective picture.

Once a structure for the action is chosen, the path integral for the theory can be formally defined. Subsequently, the complete (although formal) quantum dynamics can be fully specified by deriving the Schwinger–Dyson equations for $n-$point correlation functions from the path integral formalism.
In quantum field theory, the Schwinger–Dyson equations are one way to organize and sum in a natural way the infinitely many diagrams that contribute to $n-$point functions \cite{ShwingerDyson}. In this sense, they automatically contain non-perturbative information and encode the complete quantum dynamics of the GFT models \cite{ShwingerDysonFieldTheory}. 
The equivalence to other non-perturbative methods such as the Bogoliubov canonical transformation and the Gaussian variational Ansatz, is well established. 
In the continuum limit of GFT models it is expected that the Schwinger–Dyson equations would admit an interpretation as Hamiltonian and diffeomorphism constraints of the quantum gravity theory, and thus provide the definition of the physical inner product (at least in a regime where topology changes are suppressed) for a non-perturbative domain. 

The same quantum dynamics can also be given in an operator form for a GFT model with an action $S[\varphi,\bar\varphi]$ given by \eqref{Action}. The quantum equations of motion for a generic state $\ket{\Psi}$ can be simply written as
\begin{equation}\label{QuantEoM}
	\frac{\delta\hat{S}[\varphi,\bar\varphi]}{\delta\bar\varphi(g_I)}\ket{\Psi}=0\ ,\end{equation}
together with a second equation, obtained from the variation of the action with respect to $\varphi(g_I,\phi^J)$.
As mentioned above, the kinetic ${\cal K}$ and interaction ${\cal V}$ kernels in the generic action \eqref{Action} can be chosen so as to reproduce the edge and vertex amplitudes of a spin foam model; i.e. a perturbative expansion of the GFT partition function around the Fock vacuum can be made to coincide with the expansion of the spin foam model.
Nonetheless, when dealing with interacting field theories, solutions to the equation cannot be easily obtained; in fact, a general solution is not known without further approximations\footnote{The kinematical Fock space is known to be troublesome for defining an interacting quantum field theory.
According to Haag’s theorem one should not expect solutions to the quantum dynamics to be defined as elements on the Fock space.}.

The GFT approach deals with problems that are analogous in condensed matter physics and hence it made use of its ideas and methods. It proceeds by seeking for some condensate state that can play the role of a new, nonperturbative vacuum of the theory and that approximates the full solution state $\ket{\Psi}$, at least for a restricted set of cosmological observables without making use of an effective Wheeler–DeWitt equation \cite{QCfromQGCLattice}.
In the next section it will be shown that when restricting to states represented by a wavefunction obeying the homogeneity principle, consistent effective quantum homogeneous cosmologies are obtained \cite{HomogeneousCosmologies,HomogeneousWaveFunction,NotSU2Group}.

\section{GFT condensed phase}\label{CondensatePhGFT}

The program of GFT condensate cosmology has been proposed in \cite{FriedmannEmergent,NotSU2Group,HomogeneousCosmologies} (see Refs. \cite{ReviewCosmology,UniverseAsQGC,Pithis.Sakellariadou.Review} for reviews on the topic and their application to homogeneous cosmology). Instead of employing any symmetry reduction, the guiding principle for its construction steam on the analogy with condensed matter theory and the physics of many-body quantum systems. In this picture cosmology is represented as the hydrodynamic approximation of the quantum gravity theory.
The main purpose tackles the identification of quantum states in a full quantum field theory for $\varphi$ (function of four arguments valued in the SU(2) and minimally coupled to at least four scalar matter fields degrees of freedom\footnote{For a choice that can be generalized \cite{NotSU2Group,HomogeneousCosmologies}.}) and its possible consistent interpretation as a continuum geometry representing a macroscopic and nearly homogeneous description which can be associated to cosmological spaces.
This restricted class of states are ruled by an effective dynamics which can be understood as the cosmological evolution and must be extracted out of the fundamental quantum dynamics.

The Fock space construction previously presented proves technically very useful in order to address the problem of how to extract continuum physics from GFT.
In the full theory, fundamental quanta are associated to the $3-$simplices (alternatively spin network-type degrees of freedom) presented in the previous section that can be seen as building blocks of three-dimensional simplicial geometries representing the (boundary) states of the theory, with their quantum simplicial geometric properties encoded in the group-theoretic data; equivalently, the dual picture permits to associate the perturbative expansion of the $n-$point functions which produces a sum over Feynman diagrams, which are not graphs but $4-$dimensional cellular complexes, weighted by a discrete gravity path integral with the same group-theoretic data as dynamical variables.
The states of interest are required to have a very large (infinite on the approximation of interest) number of excitations over the Fock vacuum of this field. These microscopic degrees of freedom need to be coarse grained to obtain an elementary piece of continuum space that could serve as the starting point for GFT hydrodynamics.
In this spirit, the hydrodynamic approximation implies the negation of the microscopic dynamics (which can only be consistently determined by solving all the Schwinger-Dyson equations of the theory) in pursuit of global ones, which encodes the information corresponding to the phase space of homogeneous geometries in terms of a suitable probability distribution.
The hypothesis of condensation hence determines the form of GFT hydrodynamics. The rigorous coarse graining procedure in the full theory implies to capture the collective behaviour by a single statistical distribution over the space of ‘single-particle’ data \cite{BEC}. In this manner, the theory reflects the distinction between the continuum and the semi-classical approximation, which we have emphasized in the Introduction.

The GFT framework suggests that continuum physics is nothing more than one of the (possibly many) macroscopic phases in which the same quanta of the theory may be organized; one phase in which space and time notions emerge from the collective behaviour of discrete `geometries' through a phase transition.
Various indications for such transitions have been found, but a complete understanding of such phenomenon remains open. Simple models can be found in the context of functional renormalization group analysis in Refs. \cite{CondensatePhaseWithSymmetryBreaking}, or using Landau–Ginzburg theory in \cite{LandauGinzburgFormulation}.

\subsection{The mean field approximation}

The construction of collective states steam on the condensate hypothesis, which suggests that each fundamental $3-$simplex (or spin network vertex) has the same information, at least in its first approximation.
One is able to construct many different condensate states under this hypothesis, each of them characterized on how their constituents are glued to each other.
The simplest application of this criteria steam in the analogies with the Gross-Pitaevskii ansatz for Bose--Einstein condensates \cite{GP}; apart from choosing states satisfying the wavefunction homogeneity, one also neglects the connectivity information between their constituents. This connectivity can be understood in terms of entanglement and correlations between the quanta of the condensate, which in this simplest approximation vanishes.
These states are coherent states of the GFT field operator corresponding to an infinite superposition of states describing disconnected tetrahedra and characterized by a macroscopic occupation number. This `single-particle' condensate state is written as
\begin{equation}\label{WaveFunctionCondensate}
	\ket{\sigma}\equiv {\cal N}(\sigma)\exp\big(\,\hat\sigma\,\big)\ket{0}\ ;\end{equation}
where the normalization factor is given by
\begin{equation}
	{\cal N}(\sigma)\equiv\exp\Bigl[\,-\tfrac12\int\text{d}\phi^J\ \bigl|\sigma(\phi^J)\bigr|^2\,\Bigr]\end{equation}
and the condensate operator is defined as
\begin{equation}\label{Opperators}
	\hat\sigma\equiv\int\text{d}^4g\,\text{d}^4\phi\ \sigma(g_I,\phi^J)\,\hat\varphi^\dagger(g_I,\phi^J)\ .\end{equation}
$\sigma(g_I,\phi^J)$ describes a single collective function which is analogous to the order parameter in the context of the Gross-Pitaevskii model.
This state has an important property: it is field coherent since it is an eigenstate of the field annihilation operator $\hat\varphi$, i.e.
\begin{equation}
	\hat\varphi(g_I,\phi^J)\ket{\sigma}=\sigma(g_I,\phi^J)\ket{\sigma}\ ,\end{equation}
which is at the heart of its use as a robust and classical-like quantum state. As a general definition of such state, it acquires a nonvanishing expectation value: 
\begin{equation}
	\sigma(g_I,\phi^J):=\bra{\sigma}\hat\varphi(g_I,\phi^J)\ket{\sigma}\neq0\end{equation}
which is clearly different from the Fock vacuum in which $\bra{0}\hat\varphi(g_I,\phi^J)\ket{0}=0$. This approximation is a drastic one, but as it is useful for modeling weakly interacting Bose--Einstein condensates, it will be quite reasonable for constructing a spatially flat FLRW spacetime. In fact, the state \eqref{WaveFunctionCondensate} satisfies the homogeneity property, which requires for the state to be determined by a single-particle wavefunction.
Under this simple approximation, correlations between GFT quanta are ignored, and in the mean field-approximation where no fluctuations are present, $\sigma(g_I,\phi^J)$ represents a homogeneous condensate wave-function that can be directly understood as a classical GFT configuration. These geometric states together with the attached fields $\phi^J$ can macroscopically distinguish between different points over the condensate. Therefore, on the emerged spacetime one is able to define a complete relational dynamics. 
In this sense, the quantized space does not reside ‘‘somewhere’’, but it itself defines the {\it locus}; yet the hydrodynamic limit of GFT defines this {\it locus} as an effective theory that is not useful for describing the dynamics of the quanta, but only the collective behaviour of a big number of them. Clearly, fluctuations over this condensed phase which have been ignored can (and should) be re-established as a natural feature due to quantum uncertainty.

In the second-quantized framework, the simplest one-body observable that can be constructed is the number operator
\begin{equation}\label{NumberOp}
	\hat N(\phi^J)=\int\text{d}^4g\,\hat\varphi^\dagger(g_I,\phi^J) \,\hat\varphi(g_I,\phi^J)\,. \end{equation}
The wavefunction is not normalized; rather its norm determines the number of uncorrelated quanta in the given state
\begin{equation}\label{NumberOPMeanField}
	N(\phi^J)=\bra{\sigma}\hat{N}(\phi^J)\ket{\sigma}=\int\text{d}^4g\,\bigl|\sigma(g_I,\phi^J)\bigr|^2\quad(<\infty)\end{equation}
which is the (finite) expectation value of the operator \eqref{NumberOp} at a value $\phi^J$ for each of the four fields.
The homogeneity condition for the wavefunction can be defined over more general condensates containing additional topological structure and defined by a sum over connected graphs of arbitrary complexity \cite{HomogeneousWaveFunction}.

In the remaining part of this section, the two main testable structural consequences of the cosmological principle, namely homogeneity and isotropy, will be discussed, with emphasis on how they are both implemented in GFT. Indeed, the GFT condensate approach involves cosmological models which reproduce the Friedmann–Lema\^itre–Robertson– Walker (FLRW) spacetime, but in a semiclassical limit where the field operators are replaced by classical fields. This approximation seems to be suitable enough for describing the emergence of a macroscopic and nearly homogeneous universe, where small spatial gradients on the effective geometry can take place. In complete analogy with condense matter theory, if interactions get stronger, the mean-field approximation breaks down. This behaviour is probably expected in the early universe or near the centres of black holes where quantum effects are relevant because of high curvatures; a nice discussion on the ranges of validity of the mean-field approximation can be found in \cite{FriedmannEmergent}.

The Hartree–Fock mean field approximation \eqref{WaveFunctionCondensate} is the most simple collective wavefunction constructed under the homogeneity principle.
Its dynamics can be looked for in the expectation value of the (normal ordered) operator equations of motion obtained from \eqref{QuantEoM}. However, one can appeal to the Schwinger–Dyson equations to be solved approximately
\begin{equation}\label{EoM}
	\bra{\sigma}\frac{\widehat{\delta S}}{\delta\bar\varphi}\ket{ \sigma}=\frac{\delta S[\sigma,\bar\sigma]}{\delta\bar\sigma( g_I,\phi^J)}=-{\cal K}\,\sigma(g_I,\phi^J)+\frac{\delta{\cal V} [\sigma,\bar\sigma]}{\delta\bar\sigma(g_I,\phi^J)}=0\ .\end{equation}
This expression provides a quantum cosmology-like equation for the ‘wavefunction’ $\sigma$ (similar to those obtained in Ref. \cite{NonLinearLQC}).
The underpinning of the approximation into coherent states is only valid in regimes where the interaction term, which contributes with a non-linear term of the order of $\bar\sigma^4$, is subdominant. However, at some point interactions should become dominant since the particle number scales as $|\sigma(g_I,\phi^J)|^2$ and the potential ${\cal V}$ contains higher powers of $\sigma$ and $\bar\sigma$. These aspects will become relevant for our discussion in Section \ref{Rods}.
In this context, the microscopic dynamics of the GFT quanta can be hydrodynamically described in terms of the collective variable $\sigma(g_I,\phi^J)$, which is the condensate wavefunction.

This hydrodynamic limit is understood as an effective collapse of the Schwinger–Dyson tower of equations into the simplest one. The simplicity of the state $\ket{\sigma}$ makes this equation to coincide with the condensate wavefunction, which is the one-particle correlation function. 
The continuum nature of the picture arises from the fact that, given the equivalence with the path integral formulation, the coherent state is given by an infinite sum over numbers of disconnected spin networks nodes (implicitly a sum over ``not yet connected" graphs). Therefore $\ket{\sigma}$ is a non-perturbative state with respect to the Fock vacuum, but now playing the role of a new non-perturbative vacuum of an effective theory obtained after the hydrodynamical limit. 

Analogously to the Gross–Pitaevskii equation, the dynamics of the condensate described by \eqref{EoM} is nonlinear, as is expected in the hydrodynamic context, while all dynamical equations on the Hilbert space and on the GFT Fock space remain linear.
This equation is of course a weaker condition than \eqref{QuantEoM}; in terms of the truncated Schwinger–Dyson equation, the theoretical error in the resulting effective theory can be estimated by the magnitude of the neglected terms. These terms can be reconstructed in terms of the non-Fock representations for describing interacting fields. With this inequivalent representation of the canonical commutation relations (with respect to the free theory) the interacting theory naturally provides fluctuations over the homogeneous background associated with inhomogeneities of the condensate.

To get an initial insight on the effective dynamics of GFT condensates, two approximations are usually imposed. The first one involves all symmetries of the employed free massless scalar fields $\phi^J$ used as matter to introduce relational cosmological observables \cite{FriedmannEmergent}. 
As mentioned, these matter reference frames allow us to define an effective dynamics formulated exclusively in relational terms, where spacetime points in the emergent spacetime description can be distinguished\footnote{See Section \ref{HomogeneousUniverse} for clocks and Section \ref{Rods} for rods.}. 
Concerning the symmetries of the material clocks and rods, the GFT dynamics should be invariant under
\begin{enumerate}
\item constant (arbitrary) shifts $\phi^J\mapsto\phi^J+\phi^J_0$ ,
\item the time-reversal or parity transformation $\phi^J\rightarrow-\phi^J$ ,
\item rotations $\phi^i\rightarrow O^i{}_j\,\phi^j$, with $O^i{}_j\in$ O(3) and $i,j=1,2,3$ .\end{enumerate}
the first symmetry forbids any explicit dependence on $\phi^J$. 

As stated before, GFT condensate cosmology has been applied to a general class of GFT models without specifying the forms of ${\cal K}$ and ${\cal V}$. However, when regarding cosmological applications, these class of models are usually employed in an effective field theory (or hydrodynamic) approximation; see for instance Refs. \cite{FriedmannEmergent,BouncingCosmologies,CosmologicalPerturbations,ScalarField,InhomogeneousUniverse}.
Assuming that the GFT action is invariant under the three symmetries $1-3$ of the free and massless scalar fields given above, the expansion for ${\cal K}$  in derivatives with respect to $\phi^J$ is forced to be 
\begin{equation}\label{KineticTermExpanded}
	{\cal K}={\cal K}^0+\widetilde{\cal K}^1\frac{\partial^2}{\partial(\phi^0)^2}+{\cal K}^1\sum_{i=1}^{3}\frac{\partial^2}{\partial(\phi^i)^2}+\dots\end{equation}
The dots stand for higher derivatives which are suppressed. Let us note that the coefficients ${\cal K}^i$ are still differential operators with respect to the SU(2) variables $g_I$, but with no explicit dependence on $\phi^J$.
In this way, a truncation up to second order derivatives provides a low-energy GFT dynamics that would agree with cosmology on large scales, and the truncation up to second order derivatives \eqref{KineticTermExpanded} is a good approximation to the full kinetic term.
Under this approximation, one assumes that the fluid density $\varphi(\phi^J)$ varies slowly with respect to its arguments, particularly with respect to the relational time. This is what allows the suppression of higher order derivatives with respect to the scalar field variable in the Taylor expansion. Nonetheless, at it will be shown in Section \ref{FrdmnUniv}, this condition may not be satisfied by the condensate wavefunction representing the cosmological spacetime. As explicitly shown in \cite{FriedmannEmergent}, for satisfying the Friedmann equation at late clock times $\phi^0$, it is required an exponential behaviour for the volume operator, and thus for the wavefunction as well. Progress in this issue have been recently discussed in \cite{MarchettiOriti2020}.

The second approximation usually made for extracting physical states of the theory is to consider for the building blocks of geometry to be all in the same microscopic configuration and all in a weakly interacting regime in which the effect of ${\cal V}$ on the dynamics can be neglected. This implies for the GFT quanta to be uncorrelated; necessary condition for defining the coherent state $\ket{\sigma}$ in a mean-field treatment given by Eq. \eqref{WaveFunctionCondensate}.
This drastic approximation is not strictly necessary. In fact, it is not suitable for strong coupling regimes and it breaks down with the grown of the particle number intervening in the picture. The free approximation is valid only in a mesoscopic regime where the particle number for a given volume of the state is not so large. Some studies include the potential ${\cal V}$ of the effective dynamics for some particular models of GFT condensates\cite{Pithis.Sakellariadou.Tomov2016}. In some cases, interaction terms become important at late times, after a prolonged phase of acceleration, and they can even lead to a recollapse of the universe, while preserving the bounce that replaced the initial singularity \cite{GFTInteracting}.

Under both approximations, replacing the general expansion for the kinetic kernel \eqref{KineticTermExpanded} in the r.h.s. of Eq. \eqref{EoM}, one gets the following equation of motion
\begin{equation}\label{EoMGP}
	\biggl({\cal K}^0+\widetilde{\cal K}^1\frac{\partial^2}{\partial (\phi^0)^2}+{\cal K}^1\sum_{i=1}^{3}\frac{\partial^2}{\partial (\phi^i)^2}+\dots\biggr)\sigma(g_I,\phi^J)=0\,. \end{equation}
where the dependence on quantum geometric data is encoded in the terms ${\cal K}^0$ and ${\cal K}^1$ which are functions of the group elements (and/or their derivatives).
At a quantum level, one approaches this equation using Peter–Weyl theorem \cite{PeterWeyl} to decompose the wavefunction into SU(2) representations. The left ``gauge symmetry" in \eqref{GaugeLeftInvariance} implies for the GFT condensate to obey the identity $\sigma(g_I,\phi^J)=\sigma(hg_I, \phi^J)$ for all $h\in$ SU(2). It is desirable to give a precise geometric interpretation to the condensate as a continuous and homogeneous spatial geometry. In this picture, the condensate wavefunction is interpreted as a probability distribution on the space of such homogeneous geometries. However this interpretation requires a right invariance under the diagonal group action for the condensate; i.e. $\sigma(g_I,\phi^J)=\sigma(g_Ik,\phi^J)$ for all $k\in$ SU(2). This yields for the state to only contain the gauge-invariant degrees of freedom of a tetrahedron. Consequently $\sigma$ becomes a function on SU(2)$\setminus$SU(2)$^4$/SU(2), which is isomorphic to the space of connection degrees of freedom of a homogeneous universe in LQC \cite{LQGHomogeneousUniverse}. However, this symmetry is not one of the GFT field, as it is the left invariance\footnote{This choice is a convention and can be exchanged; on the other way around, one can start with a right ``gauge symmetry" on the $\varphi$ field and then impose the left invariance over $\sigma$ obtaining equivalent results.}, but an imposed property on certain states aiming at reducing the number of dynamical degrees of freedom.

In the mean-field approximation, self-consistency implies the existence of a regime where the field equation \eqref{EoMGP} is approximately solved only considering the first kinetic term.
The uncorrelated state solving this equation has interesting cosmological applications to be reviewed in what follows. Being a many-particle state with an analogous hydrodynamical treatment, it can contain information about the connection and the metric at many different points in space. To distinguish them, the introduced massless scalar fields represent relational clocks $\phi^0$ and relational rods $\phi^i$ defined over the condensate. Apart from the symmetries 1–3, no assumptions have been made over these fields.

\subsection{Isotropic condensed states}\label{Isotropic}

At this point we have obtained a condensate-like structure characterized by a nearly spatial homogeneity; however, one more feature is needed for making contact with usual cosmology; namely, isotropy.
In what follows, a further restriction on the structure of the wavefunction of the condensate will be imposed. As it will be shown in the following and in Section \ref{FrdmnUniv}, the restriction to isotropic modes for the microscopic states leads to further simplifications that allow us to reconstruct isotropic quantities like the spatial volume, the cosmological scale factor and thus the Hubble rate from $\sigma(g_I,\phi^J)$. 
In the GFT context, it is argued that the natural way to require isotropy is to impose the most ‘isotropic’ condensate configuration. In classical geometry one would think of equilateral tetrahedra whose four faces are equal and the resulting volume is maximized. This ideas have been translated to the quantum picture in \cite{FriedmannEmergent,BouncingCosmologies}.

Regarding the analogy with equilateral tetrahedra, one can postulate that each of the four links/faces $g_I$ are coloured with spin$-j$ irreducible representation of SU(2). The correspondence between the equal areas of the faces of the tetrahedron and the restriction to an expansion over isotropic modes only is understood as expanding over the same four $j_I$, i.e. $j_1=j_2=j_3=j_4$, one for each (of the same) coloured link associated to each node of the spin network. In this approximation, the spin network vertices are said to be monochromatic and, together with the homogeneous restriction, all of them are exactly equal. Secondly, with respect to the volume maximization of the classical tetrahedron, the SU(2)$-$invariant subspace reduces the wavefunction to a simple form if written in terms of linear combinations of a pair of suitable intertwiners: $\bar{\cal I}^{j,\imath_l}_{mn}$ and ${\cal I}^{j,\imath_r}_{mn}$, one associated to the left gauge invariance and one to the right closure condition. These intertwiners define invariant mappings between SU(2) representations and they are elements of the Hilbert space of states of a single tetrahedron. The intertwiners should be chosen so that they are eigenstate of the LQG volume operator with an associated eigenvalue being the largest possible for the given $j$.

By decomposing the wavefunction $\sigma(g_I,\phi^J)$, solution of the equation of motion \eqref{EoMGP}, into a basis of orthonormal functions given by the Wigner $D_{mn}^j(g_I)-$matrices, we group all dependence on $g_I$ in fixed functions $\mathbf{D}^j(g_I)$, which are an appropriate convolution of four Wigner $D-$matrices with intertwiners\footnote{In the convolution of Wigner $D-$matrices with SU(2) intertwiners, the usual range of values for the magnetic indices is taken: $-j\leq m$, $n\leq j$. The indices $\imath$ labels the possible intertwiners elements in a basis of the Hilbert space; $\imath_l$ and $\imath_r$ points to the imposition of the left and right invariance to the field. To a detailed construction of the wavefunction see for instance \cite{FriedmannEmergent}; here we just sum up the main steps for deriving a cosmological sector from the full theory.}, such that quantum geometric properties of the field are stored in the scalar wavefunctions $\sigma_{j_I,\,\imath_l\imath_r}\equiv\sigma_j(\phi^J)$ that now only depend on the volume of the tetrahedron (or equivalently, on the surface area of one of its faces), as well as on the scalar field $\phi^J$. Therefore, the restricted mean field expanded in irreducible SU(2) representations is written as
\begin{equation}\label{WaveFunctionExpandedIsotropic}
	\sigma(g_I,\phi^J)=\sum_{j\in\tfrac{\mathbb{N}_0}2}\sigma_j(\phi^J)\mathbf{D}^j(g_I)\end{equation}
where the coarse-grained degrees of freedom are now captured by each of the wavefunctions $\sigma_j(\phi^J)$.
Refs. \cite{NonIsotropic,NonIsotropic2} discuss whether the restriction to expansions in only a single spin $j$ labelling the irreducible representations of SU(2) can be relaxed, together with their consequent effective dynamics in the large-scale limit. Recall that lifting the isotropic restriction allows to investigate anisotropic GFT condensate configurations. According to \cite{NonIsotropic2}, anisotropies play an important role only at small values of the relational clock $\phi^0$ (i.e. at small volumes), whereas at late times the isotropic mode become dominant.

For the usual GFT actions, the kinetic operator ${\cal K}$ only contains derivatives, but no explicit dependence on $g_I$. In the common situation where all terms in the expansion of ${\cal K}$ in Eq. \eqref{EoMGP} are general functions of the Laplace–Beltrami operators with respect to the SU(2) variables $g_I$, we can define the following coefficients
\begin{equation}
	{\cal K}^0{\bf D}^j(g_I):=-B_j{\bf D}^j(g_I)\ ,\qquad \widetilde{\cal K}^1{\bf D}^j(g_I):=A_j{\bf D}^j(g_I)\ ,\qquad {\cal K}^1{\bf D}^j(g_I):=C_j{\bf D}^j(g_i)\ ; \end{equation}
$A_j$, $B_j$ and $C_j$ are $j-$dependent couplings depending on the original GFT kinetic terms and with no further derivatives. Each Laplacian acting on each $g_I$ contributes with an eigenvalue $-j(j+1)$. Recall that ${\bf D}^j(g_I)$ encode the monochromatic (equilateral) character of the spin network nodes (tetrahedra).
The Wigner matrices are eigenfunctions of the SU(2) Laplacian, then the Peter–Weyl decomposition leads to a decoupling of \eqref{EoMGP} into independent equations for each $j$, written as
\begin{equation}\label{EoMIsotropic}
	\biggl(-B_j+A_j\,\partial^2_{\phi^0}+C_j\sum_{i=1}^3\partial^2_{\phi^i}\biggr)\sigma_j(\phi^J)=0\ .\end{equation}
In homogeneous configurations, the handling of ``rods" $\phi^i$ loses meaning. Therefore, when deriving the global aspects of a FLRW cosmology only the first two terms matter. In the next section, it will be shown that the two corresponding coefficients, $A_j$ and $B_j$, can be constrained when requiring the theory to be compatible with Friedmann equations. However, condensate fluctuations are expected to break the homogeneity; analogously as when in the cosmological model one considers deviations from the cosmological principle.
In such a case, ``rods" must be reintroduced to locate these deviations, whose power spectrum is probably expected to be associated to classical inhomogeneities observed in the cosmic microwave background (CMB) spectrum.
At this point, the third term becomes meaningful. We will return to discuss this topic in Section \ref{Rods}. 

Interestingly, if one expands $\sigma_j$ in Fourier modes with respect to the spatial coordinates pictured as the scalar fields $\phi^i$, a complete set of solutions to \eqref{EoMIsotropic} can be obtained; this is
\begin{equation}\label{Solution}
	\sigma^{K_i}_j(\phi^J)=e^{iK_i\phi^i}\left[\alpha^+_j\exp\left(\sqrt{\tfrac{B_j+C_jK^2}{A_j}}\ \phi^0\right)+\alpha^-_j\exp \left(-\sqrt{\tfrac{B_j+C_jK^2}{A_j}}\ \phi^0\right)\right], \end{equation}
with $\alpha^+_j $ and $\alpha^-_j $ as arbitrary constants. The coupled scalar fields act as `tools' to define local coordinates; hence any coordinate system constituted by physical degrees of freedom must be relational. In the limit in which they are turned off, we obtain a homogeneous solution. Being the condensate wavefunction at a given time $\phi^0$ entirely determined by just one geometric quantity, the spin $j$, the only geometric quantities that can be extracted from this condensate wavefunction are isotropic quantities like the total spatial volume, the Hubble rate, etc.

\section{The Friedmann universe}\label{FrdmnUniv}

Let us now consider proper GFT cosmological models. In this section it will be shown how to obtain a FLRW universe from the condensate wavefunction \eqref{Solution}. However, we first give the main ingredients of cosmological implications derived from General Relativity to show explicitly how the previous approximations to GFT quantum gravity formalism lead us to a quantum picture consistent with classical results in the continuum and semiclassical limit.

\subsection{Classical general relativity}\label{ClassicalGR}

It is well known how to introduce physical reference frames and how to define relational dynamics in general relativity. Let us consider a massless free scalar field that plays the role of a relational clock in a flat FLRW metric of the form
\begin{equation}
\text{d}s^2=-N^2(t)\ \text{d}t^2+a^2(t)\ \text{d}x_3{}^2\ . \end{equation}
The structure of the metric entails a foliation for the universe on isotropic and homogeneous hypersurfaces $\text{d}x_3{}^2$ with flat intrinsic geometry in $\mathbb{R}^3$ parametrized, relative to one another, by a scale factor $a(t)$. 

The action for the scalar field, considering limiting cases in which the backreaction of reference matter on the geometry can be neglected, is
\begin{equation}
	S_\phi=-\frac{1}{2}\int\text{d}^4x\sqrt{-g}\ g^{\mu\nu}\partial_\mu \phi\,\partial_\nu\phi\ ;\end{equation}
hence the matter clock obeys the Klein–Gordon equation, which reduces to
\begin{equation}
	\nabla^\mu\nabla_\mu\phi=0\qquad\Rightarrow\qquad\frac{\text{d}}{\text{d}t}\left(\frac{a^3}{N}\frac{\text{d}\phi}{\text{d}t}\right)=0\qquad\Rightarrow\qquad\frac{a^3}{N}\frac{\text{d}\phi}{\text{d}t}=\text{constant}\,.\end{equation}
The assumption of a nonnull constant provides a good characteristic for the clock $\phi$; as its corresponding momentum $\pi_\phi$ is conserved, it has a monotonic evolution and thus, it can be written as $\phi=\phi_0\,T$. If for instance $\phi_0$ has dimensions of mass, the `temporal' scalar field $T$ becomes dimensionless. 

The other equation to solve is
\begin{equation}
	\left(\frac{\text{d}a}{\text{d}T}\right)^2= \frac{4\pi G}{3}\ \phi_0{}^2\ a^2\,, \end{equation}
which is a Friedmann equation giving two independent solutions:
\begin{equation}\label{FLRWSolution}
	a(T)=a_0\exp\left(\pm\sqrt{\frac{4\pi G}{3}}\ \phi\right) ,\end{equation}
each of them corresponds to an expanding or contracting universe, respectively. According to the classical solution, a singularity $V\rightarrow0$ appears in the far past where, due to our choice of time field $\phi$, corresponds to infinity. However, this singularity is reached in a finite proper time, if written in the propitious coordinates.

A quantum theory of gravity coupled to a massless scalar field should reproduce at some point the last equation but avoiding the singularity behaviour. This is the basic idea since the early days of quantum cosmology \cite{CosmologyWithScalarField} that later on also informed the foundations of LQC. The requirement for the temporal coordinate to satisfy the harmonic condition avoids quantization ambiguities when choosing the lapse function \cite{LQCFoundations}.

\subsection{The homogeneous universe in GFT}\label{HomogeneousUniverse}

At this point we have a condensate with an isotropic structure imposed from the quantum symmetries. If spatial homogeneity is also desired, this would correspond to demand the Fourier mode $\vec{K}=0$ for the mean field solution in the general solution \eqref{Solution}. 
This requirement makes the rods $\phi^i$, with $i=1,2,3$, meaningless, as there is no need to refer to locations over an exactly homogeneous state. Then, the mean field would have the form
\begin{equation}\label{WaveFunctionHomogeneous}
	\sigma_j (\phi^J )\equiv \sigma_j(\phi^0)\ ,\end{equation}
being only a function of one scalar “time field” $\phi^0$, playing the role of a relational clock. 
This implies for the condensate wavefunction to become
\begin{equation}\label{SolutionHomogeneous}
	\sigma_j (\phi^0 ) = \alpha^+_j \exp\left(\sqrt{\frac{B_j}{A_j}}\phi^0\right)+ \alpha^-_j \exp\left(-\sqrt{\frac{B_j}{A_j}}\phi^0\right).\end{equation} 
If we assume that the condensate mean field takes its homogeneous form, the associated (background) universe of course would result homogeneous.

Once the mean field solution is found, it is of interest to define the relational $3-$volume for this state. This can be done by means of the second-quantized vertex volume operator. This one-body operator generically would define for the mean field wavefunction \eqref{Solution} a local volume element at the spacetime point specified by values of the reference fields, this is $\hat V(\varphi^J)$. In the particular case of the homogeneous universe under consideration, the constrain \eqref{WaveFunctionHomogeneous} would then define the element volume only at a given relational ``time" $\phi^0$
\begin{equation}\label{Volume}
	\hat V(\phi^0)=\int_{\text{SU(2)}^4\times \text{SU(2)}^4} \text{d}^4g\,\text{d}^4g' \, \hat\varphi^\dagger(g_I,\phi^0)V^{LQG}(g_I,g'_I) \hat\varphi(g'_I,\phi^0)\ .\end{equation}
The matrix elements $V^{LQG}(g,g')\equiv\bra{g_I}V^{LQG}\ket{g'_I}$ are the matrix element of the volume operator between single-vertex spin networks states in LQG  \cite{GeometryEigenvalues}.
Although there are several different definitions of the volume operator in the theory \cite{VolumeOpLQGGeneric}, all of them agree when $4-$valent vertex are considered \cite{VolumeOpLQG}. 
In fact, it is helpful to choose a basis of intertwiners ${\cal I}$ that diagonalizes the action of the LQG volume operator on a spin-network node. Hence, the heuristic picture for spin networks implies for the underlying quantum theory that the quanta $\varphi$ will carry a definite volume given by the corresponding eigenvalue of the LQG volume operator, therefore interpreted as “grains of space”.

This is not the case for the collective wavefunction. Let us now go back to the homogeneous GFT state $\sigma_j(\phi^0)$, the expectation value for the volume operator \eqref{Volume} can be evaluated immediately when coherent states of the form \eqref{WaveFunctionCondensate} are considered
\begin{equation}
\langle\hat{V}(\phi^0)\rangle=\int\text{d}^4g\,\text{d}^4g'\ \bar\sigma(g_I,\phi^0)\,V(g_I,g'_I)\,\sigma(g'_I,\phi^0)\ .\end{equation}
This result corresponds to the total $3-$volume at a relational time $\phi^0$, associated to such condensate state. 
This procedure is not a novelty of GFT; for instance, the total volume of the universe at a fixed value of the scalar field is one of the main relational observables of interest in LQC \cite{LQCFoundations,LCQReview}

If we also impose the isotropic wavefunction constraint discussed in \eqref{WaveFunctionExpandedIsotropic}, since the volume operator is diagonal when written in terms of SU(2) representations, the volume expectation value of the condensate in such a state reduces simply to
\begin{equation}\label{VolumeBackground}
	\langle\hat V(\phi^0)\rangle=\sum_{j=0}^\infty V_j\, \bigl|\sigma_j(\phi^0)\bigr|^2\ .\end{equation}
The latter expression is written in terms of the local particle number density for each quanta of spin $j$. The approximate eigenvalue of the first quantized volume operator acting upon a node, although depending on the intertwiner used to define ${\bf D}_j(g_I)$, is very well approximated for each $j$ by $V_j\sim \ell_{Pl}^3\,j^{3/2}$.

The evolution of the local volume elements then provides the macroscopic behaviour of the state \eqref{SolutionHomogeneous}, which will depend on the choice of the initial parameters $\alpha_j^+$ and $\alpha_j^-$. 
Some general statements can be sketched out for some GFT models: if the ratio $B_{j_0}/A_{j_0}$ is positive and develops a maximum for a given $j=j_0$, except for the fine-tuned cases with $\alpha_j^+=0$ or $\alpha_j^-=0$, the spin $j_0$ will dominate over all others. Hence for almost any condensate homogeneous wavefunction of the form \eqref{WaveFunctionHomogeneous}, its associated volume will asymptotically behaves as
\begin{equation}\label{VolumeAsymptotic}\begin{split}
	\langle\hat{V}(\phi^0)\rangle\hspace{7pt}\xrightarrow{\ \phi^0 \rightarrow-\infty\ }\hspace{9pt}\bigr|\sigma^-_{j_0}\bigr|^2 \exp\left(-2\sqrt{\frac{B_{j_0}}{A_{j_0}}}\ \phi^0\right)\ ,\\[0.5ex]
	\langle\hat{V}(\phi^0)\rangle\hspace{7pt}\xrightarrow{\ \phi^0 \rightarrow+\infty\ }\hspace{9pt}\bigr|\sigma^+_{j_0}\bigr|^2 \exp\left(+2\sqrt{\frac{B_{j_0}}{A_{j_0}}}\ \phi^0\right)\ ;\end{split}\end{equation}
where the global constants are related to the volume eigenvalue assigned to the spin $j_0$ by $|\sigma^\pm_{j_0}|^2=V_{j_0}|\alpha^\pm_{j_0}|^2$.
In such a situation an exponentially large number of quanta are characterised by a single spin $j_0$ excitation, implying mainly a constant volume per quantum.
This domination of a single and small spin component in the cosmological dynamics of the homogeneous and isotropic background can be shown to take place at later times \cite{LowSpin}; however it is also achieved exponentially fast and hence it can be expected to be an acceptable approximation also at earlier times \cite{NonIsotropic}.
Thus, the evolution of the total volume only depends on the growth of the number of particles with spin $j_0$ given by the exponential factor in Eq. \eqref{VolumeAsymptotic}.

These GFT states closely match the heuristic relation between LQG and LQC, where this type of quantum states are usually assumed \cite{LQGLQC}.
Despite the exact relation between both theories remains open, some proposals analyse a cosmological sector of LQG built up on states with large number of spin network nodes, all labelled by the same quantum numbers. The nodes are considered to be disconnected and all links are dressed with the same SU(2) representation label. Commonly, the spins are taken to be $j=1/2$, and homogeneity considerations justify the same number of links per node, typically chosen as $4-$valent nodes. Shortly, LQG also suggests to consider quantum geometry condensates where all its constituents are quanta in the same state \cite{Calcagni2014}. All these features are naturally encoded in the cosmological results obtained from the GFT formalism; hence the latter can be considered as a field theory reformulation of LQG and spin foam models. However, it is worth mentioning that a derivation of LQC from Hamiltonian formulations of LQG is a largely outstanding challenge \cite{Dapor.Liegener.Pawlowski2019}.

Interestingly, for large (positive or negative) $\phi^0$, the coefficients $A_j$ and $B_j$ are identified with the low energy (emergent) Newton constant $G$ as follows: $B_{j_0}/A_{j_0} =3\pi G$, Eq. \eqref{VolumeAsymptotic} reproduces the general solution of the Friedmann equation in \eqref{FLRWSolution}.
Therefore, the resulting expression point towards a regime where the universe expands to a macroscopic size (if $\sigma^\pm_{j_0}\neq0$). This compatibility was one of the main results of \cite{FriedmannEmergent,BouncingCosmologies} and is obtained for a range of parameters of the microscopic dynamics in a suitable semiclassical regime and, as mentioned before, for generic initial conditions \cite{LowSpin}.

It is also worth mentioning some properties characterizing this range of GFT models with the desired asymptotic behaviour. First, at small volumes, in the Planck regime, the theory interpolates between the classical expanding and contracting solutions \eqref{FLRWSolution} of the classical Friedmann dynamics. This implies that the universe undergoes a bounce, i.e. the volume elements never go through zero avoiding or `resolving' the classical big bang singularity. In fact, it is possible to show that a singularity where $\langle\hat{V}(\phi^0)\rangle$ strictly vanishes for some value of the clock field $\phi^0$ is only possible for special (hence fine-tuned) initial conditions. Therefore, instead of a singularity, there is just a very dense region where an effective quantum force appears like a repulsion that prevents the collapse.
Secondly, the asymptotic behaviour in Eq. \eqref{VolumeAsymptotic} shows an exponentially growing phase in both temporal directions: to the far past and the future (contrary to the classical solution that, as mentioned in Section \ref{ClassicalGR}, presents a singularity in the far past).
Third, properties of interest are the corrections to the classical Friedmann dynamics. Similar derivations to LQC dynamics can be found from GFT condensate cosmology; see for instance \cite{GFTCorrections}.
Indeed, when a single spin dominates, general mean field solutions provide corrections that match with the ``improved dynamics" of LQC \cite{LQCCorrections}. 

More recently, a different analysis of GFT using Hamiltonian methods has been developed in Ref. \cite{GFTHamiltonian}; this topic is further discussed in Section \ref{Closing Discussion}.
There are many results that can serve as a starting point for GFT phenomenology: the generic quantum bounce can be followed by a subsequent acceleration replacing inflation \cite{InflationWithNoInflaton}, the inclusion of interactions in the GFT \cite{NonIsotropic2,InteractionGFT} and their subsequent effects which become dominant away from the bounce.

\section{Beyond homogeneity}\label{Rods}

One of the main points of any quantum gravity theory is how to derive relevant quantities to be compared to standard cosmology and observations. An important step towards a more realistic GFT condensate cosmology program is to construct more realistic and testable situations, which would certain require to extend all results from exactly homogeneous to inhomogeneous universes and to go beyond the considered isotropic restriction. Recent advances regarding the later has already been mentioned along this text; exploration of anisotropies can be found in \cite{NonIsotropic2}.
If quantum gravity is to offer the picture of the earliest moments of our universe, anisotropic perturbations must play a role. For instance, the systematic investigation of them over an isotropic background in the vicinity of the bounce can be found in \cite{NonIsotropic}. Regarding the former, this section will briefly discuss how to address inhomogeneities and study more general configurations and their dynamics.

Modern cosmology taught us that inhomogeneities in the very early universe may be the seeds for structure formation \cite{Mukhanov}. Therefore, the extension of the GFT framework beyond the spatial homogeneity by the encompass of cosmological perturbations into the formalism, would allow to construct more realistic cosmological scenarios where inhomogeneities are present. The strategy is to compute the non-vanishing power spectrum of cosmological perturbations over the mean field state discussed in previous sections.
Following the analogy with Bose–Einstein condensates, perturbations can be added to \eqref{SolutionHomogeneous} in an analogous manner as phonons appear as deviations from condensates with exact homogeneity. In fact, GFT phonons were firstly proposed in \cite{HomogeneousCosmologies}, however their interpretation was not clear until rod matter fields were included.
With the inclusion of $\phi^i$, quantum fluctuations in the local volume could be naturally interpreted as seeds of cosmological inhomogeneities. Afterwards, these perturbations around the mean field solution are to be converted into classical inhomogeneities in a later stage of the universe \cite{InhomogeneousUniverse}. 

Different approaches have been considered to include perturbations. For instance, one can associate a constant mean field solution but only to `local' patches, labelling them by making use of the four scalar fields coordinates. The inhomogeneity relies in the fact that different patches do not necessarily have the same constant mean field solution, thus the effective homogeneous geometry does not necessary coincide among different patches. This picture for incorporating inhomogeneities is based on the so-called ‘separate universe approach’\cite{SeparateUniverse}, whose main characteristics are considered in the GFT condensate cosmology framework \cite{SeparateUniverseGFT}.

In the remaining of this section, we will focus first on quantum fluctuations of the local $3-$volume around the exactly homogeneous background condensate derived in Eq. \eqref{SolutionHomogeneous}.
With the aim of connecting GFT results to observations, since we do not have direct access to local volume densities, the cosmological perturbations of physical relevance are the matter density perturbations.
However, the relation between the later and the volume density perturbations is in general gauge dependent. We will finish this review discussing how to extend the formalism to include perturbations in the matter density.
The GFT models discussed above have enough degrees of freedom for describing inhomogeneous quantum geometries and their effective dynamics which is expected to be a more realistic picture for fundamental cosmology. Quantum fluctuations would then represent the quantum gravitational mechanism for explaining the origin of these inhomogeneities. This procedure is analogous to the usual treatment in inflation where the power spectrum of quantum fluctuations over a homogeneous quantum state is computed (instead of a quantum state on a classically perturbed geometry). These fluctuations are generically expected because of quantum uncertainty and they would {\it freeze out} producing the classical pattern on inhomogeneities currently observed in the CMB \cite{CMB}. Besides, these homogeneities would provide a lower bound on deviations from exact homogeneity in GFT \cite{CosmologicalPerturbationsInCondensates}.

\subsection{Volume density perturbations}

Let us consider the cosmological scenario in which a GFT condensate phase modeled as a background state with perturbations on top of it, evolves to the expanding universe of classical cosmology.
For introducing vacuum fluctuations, the notion of “wavenumber” is a must.
If the starting point for localizing events in time was solved by the introduction of a scalar field $\phi^0$ used as a clock to label the evolution of the geometry; the problem of localizing events in spacetime can be solved by re-establishing the four scalar fields (in four spacetime dimensions), using now these scalars as a physical coordinate system. 
The condition \eqref{WaveFunctionHomogeneous} must be relaxed so as to assign a clock and 3 rods to each point of the condensate to characterize deviations from homogeneity. Consequently, one must replace also $\phi^0\rightarrow\phi^J$ into the volume operator \eqref{Volume} from where the effective cosmological dynamics has been derived from.
Now, the operator $\hat V(\phi^J)$ refers to the differential local volume element at the spacetime location specified by the components of the $\phi^J$ field.
Scalar perturbations in cosmology are then obtained from perturbations in these local volume elements. Strictly speaking, $\hat V(\phi^J)$ corresponds to a density. The infinitesimal local volume is then $\hat V(\phi^J)\,\delta^4 \phi^J$ and the total $3-$volume is obtained by integrating over the rods $\phi^i$ at a given moment of the relational time $\phi^0$
\begin{equation}\label{VolumeIntegrated}
	\hat V(\phi^0)\equiv\int \text{d}^3\phi^i\ \hat V(\phi^0,\phi^i)\ . \end{equation}
This expression is still interpreted as the total volume of the universe modelled as a condensate state, in analogy to \eqref{Volume}.

The main idea is that cosmic structures are expected to be formed from early local volume fluctuations. In the GFT cosmology approach, this pattern is expected to be encoded in the correlation functions for the geometric observables. These correlation functions encode the `true' fundamental quantum dynamics.
For the ongoing discussion, let us compute correlations in local volume fluctuations over the state \eqref{WaveFunctionCondensate}. It is defined the local volume fluctuation operator as
\begin{equation}\label{VolumeFluctuation}
	\delta\hat V(\phi^J)=\hat V(\phi^J)-\langle\hat{V}(\phi^J)\rangle \end{equation}
with respect to the generalized volume operator $\hat V(\phi^J)$. Then, it is of interest to compute the following two-points function
\begin{equation}\label{2PointFunction}
	\langle\delta\hat{V}(\phi^J)\,\delta\hat{V}(\phi'^J)\rangle\ .\end{equation}
The idea of characterizing perturbations employing matrix elements of the one-body squared volume operator $V^2(g_I,g'_I)$ is not new.
This procedure has been first presented in \cite{CosmologicalPerturbationsInCondensates} but without referring to any notion of rods. Consequently, the results that can be derived from this formalism can only achieve global properties. Later on, in \cite{CosmologicalPerturbations} the formalism has been generalized to include rods. This modification enables us to extract local information regarding perturbations. As it will be discussed in the following, the transformation to momentum representation allows to write the power spectrum of inhomogeneities into the usual Fourier space form.

Let us now consider perturbations around exact homogeneity. These are written as
\begin{equation}\label{PerturbedState}
\sigma_j(\phi^J)=\sigma_j(\phi^0)\bigl[1+\epsilon\,\psi_j(\phi^J)\bigr]\ ;\end{equation}
where the field $\psi_j(\phi^J)$ represents condensate perturbations `located' by means of the four scalar fields.
Fourier transforming three of them, the three scalar fields corresponding to the spatial directions defined by the rods $\phi^i$ to their momenta $k_i$, notions of wave length with respect to these reference frame fields are obtained.
Following \cite{CosmologicalPerturbations}, the power spectrum for the volume perturbations can be derived computing the quantum fluctuations of the volume expressed as the two point function \eqref{2PointFunction} for the state \eqref{PerturbedState}, giving
\begin{equation}\label{VolumePerturbations}\begin{split}
	\langle\hat{V}(\phi^0,k_i)\hat{V}(\phi'{}^0,k'_i)\rangle-\langle\hat{V}(\phi^0,k_i)\rangle\langle\hat{V}(\phi'{}^0,k'_i)\rangle=\delta(\phi^0-\phi'{}^0)\sum_{j}V_j{}^2\bigl|\sigma_j(\phi^0)\bigr|^2\times\\\Bigl\{(2\pi)^3\delta^3(k_i+k'_i)+\epsilon\Bigl[\psi_j(\phi^0,k_i+k'_i)+\overline{\psi_j(\phi^0,-k_i-k'_i)}\Bigr]\Bigr\}\ .\end{split}\end{equation}
In agreement with the results obtained when considering an exact homogeneous background, the first term is naturally scale invariant with respect to the rod wavenumbers $k_i$, and its scale depends only on the reference matter through the matter clock $\phi^0$. Besides, in this very same term, the delta function in the momentum implies a deep connection between scale invariance and translational invariance. 
The second term corresponds to first deviations from the scale invariant homogeneous mean-field, associated to inhomogeneous fluctuations, which naturally have small relative amplitude.
In line with the usual cosmological perturbations, their shape must solve the condensate dynamics and they are fully determined in a two-fold manner by the coupling with the background on the one side, and by their own dynamics on the other one.

A magnitude of particular interest can be defined by the amplitude of the volume fluctuations relative to the background; this is the quotient between at least the two correlation functions \eqref{VolumePerturbations} and the squared background volume regularized by the integral over the added matter rods $\phi^i$. This background volume computed in \eqref{VolumeIntegrated} can be rewritten as 
\begin{equation}
	\langle V(\phi^0)\rangle=\int \text{d}^3\phi^i\ \sum_{j}V_j |\sigma_j(\phi^0)|^2\ ,\end{equation}
so that it is explicitly associated with the number of quanta that makes up the condensate.
Recalling the previous analysis regarding states with a dominance of a single spin $j_0$ and keeping the dominant part of  \eqref{VolumePerturbations}, the leading term of the power spectrum of such perturbations becomes
\begin{equation}\label{PowerSpectrum}\begin{split}
	{\cal P}_{\delta V}(k)=\frac{V_{j_0}{}^2\bigl|\sigma_{j_0} (\phi^0)\bigr| {}^2}{\bigl(\int \text{d}^3\phi^i\ V_{j_0}\bigl| \sigma_{j_0}(\phi^0)\bigr|^2\bigr)^2}= \frac{1}{\bigl(\int \text{d}^3\phi^i\bigr)N(\phi^0)} \ .\end{split}\end{equation}
In the last equality, the dependence with the number of quanta $N(\phi^0)=\int \text{d}^3\phi^i\ \bigl| \sigma_{j_0}(\phi^0)\bigr|^2$ has been made explicit in the denominator. Therefore, the relative amplitude of these scalar perturbations decreases as $\sim 1/N$; i.e. they decrease with the growth of the number of quanta $N$ while $\phi^0$ evolves and the universe expands. 
In the particular case in which $C_j/B_j<0$ in the equation of motion \eqref{EoMIsotropic}, at large volumes, scale invariance is approached more closely. As reported in Ref.  \cite{CosmologicalPerturbations}, inhomogeneous terms decay relative to the homogeneous background thus, under these conditions, inhomogeneous perturbations are further suppressed. Besides, if interacting GFT are considered \cite{GFTInteracting}, the obtained long-lasting accelerated expansion (after the bounce) is accompanied by a further suppression of the deviations from scale invariance, suggesting further investigations of GFT interactions.
These scaling results are in agreement with the typical relative size of fluctuations in a condensate. Analogously, these fluctuations arise naturally in the GFT condensate approach, but within a quantum gravity theory for gravity and matter, which has a properly defined ultraviolet completion.

\subsection{Matter density perturbations}

We have seen that perturbations of local volume observables around a quantum state which solves the condensate dynamics and that thus depends on the approximation scheme summarized in the previous sections, exist as quantum fluctuations given by the two-point function \eqref{2PointFunction}.
These perturbations can be expressed through the expectation values and fluctuations of local volume elements given by the GFT Fock space operator $\hat{V}(\phi^J)$ and in the hydrodynamic approximation, they have been generated with a small constant amplitude at all scales in the Fourier modes $k_i$ and their amplitudes. When only a single spin $j = j_0$ dominates, their amplitude scales as $\sqrt{V (\phi^0)}$ as one would expect in general for extensive quantities such as the volume. Consequently, its corresponding power spectrum is nearly scale-invariant; when the mean field description is exactly homogeneous, an exact scale-invariant power spectrum for volume density perturbations is found.

Although volume density perturbations are simple to compute in the full GFT formalism staying within the full quantum gravity framework (but in a hydrodynamic approximation), their relation to matter density perturbations are in general not gauge-invariant and they can not be directly related to cosmological observables.
The usual gauge freedom of cosmological perturbation theory is absent because of the use of the values of matter fields as relational coordinates. 
If the semiclassical low-curvature universe described within standard perturbation theory is preceded by a deep quantum-gravity phase in which perturbations originate as quantum fluctuations of the GFT field, this field theory must be a field theory for quantum gravity and matter.
Therefore, the full calculation must involve quantum fluctuations both in the $3-$volume element and in the matter energy density of the scalar matter. Furthermore, their relation, in general gauge dependent, must be connected to the usual gauge-invariant variables for observable scalar perturbations in cosmology.

In the following we will review how to extend the previous arguments to perturbations on the matter density.
The starting point is the total kinetic energy for the whole matter content introduced in the theory so far.
In analogy with the classical kinetic energy density of the scalar field, $\rho_{\text{kin}}=\pi_{\phi}  ^2/ (2V ^2)$, one can define its corresponding analogue in the GFT context by replacing the classical expressions with the expectation values
\begin{equation}\label{KinEnDen}
	\rho_{\text{kin}}=\sum_{I=0}^{3}\rho_{\text{kin}}^I \hspace{20pt}\text{whit} \hspace{20pt}\rho^I_{\text{kin}}=\frac{1}{2}\,\frac{\langle\hat\pi^I_{\phi} (\phi^J) \rangle^2} {\langle \hat V(\phi^J) \rangle^2}\, . \end{equation}
The self-adjoint relational momentum operator of the scalar field $\phi^I$ has been obtained in Ref. \cite{FriedmannEmergent} as the observable defined by setting ${\cal O}_{\{j_I\}}=-(i/2)\partial/\partial\phi^I$; i.e. 
\begin{equation}\label{ConjMomentum}
	\hat\pi^I_{\phi}(\phi^J)=-\frac{i}{2}\int\text{d}^4g\left[\hat\varphi ^\dagger(g_I,\phi^J)\,\frac{\partial\hat\varphi(g_I,\phi^J)}{\partial\phi^I}-\frac{\partial\hat\varphi^\dagger(g_I,\phi^J)}{\partial \phi^I}\,\hat\varphi(g_I,\phi^J) \right]\ \, , \end{equation}
and $\langle\hat V(\phi^J)\rangle^2$ is the expectation value of the local volume element \eqref{Volume}.

At leading order, perturbations in the total kinetic energy density \eqref{KinEnDen} are given by
\begin{equation}\label{Pert1stO}
	\delta\rho_{\text{kin}}(\phi^J)=\sum_{I=0}^3 \frac{\pi^I_{\phi} (\phi^J)\,\delta\pi^I_{\phi}(\phi^J)}{ V^2(\phi^J)} -2\, \rho_{\text{kin}}(\phi^J)\,\frac{\delta V(\phi^J)}{V(\phi^J)}\ . \end{equation}
As a consequence of the near-homogeneity assumption the dynamics is well approximated by ignoring spatial derivatives in favor of time derivatives.
Therefore one expects for the dynamics of the free scalar fields to be completely dominated by their kinetic energy, neglecting any gradient energy.
For a homogeneous, isotropic mean field of the form \eqref{WaveFunctionHomogeneous}, derivatives of $\sigma_j$ with respect to the rod fields vanish. Then we have
\begin{equation}
	\pi_\phi^i\equiv\langle\hat\pi^i_\phi(\phi^J) \rangle =0\,,\end{equation}
for the three spatial scalar fields, $i=1,2,3$. 
Therefore, the only nonnull conjugate momentum from Eq. \eqref{ConjMomentum} is $\pi^0_\phi(\phi^0)=(-i/2)\,[\bar\sigma(\phi^0)\sigma'(\phi^0)-\bar\sigma'(\phi^0)\sigma(\phi^0)]$.
Accordingly, Eq. \eqref{KinEnDen} simplifies to $\rho_{\text{kin}}=\rho_{\text{kin}}^0$ and thus, only the clock field $\phi^0$ contributes non-negligibly to \eqref{Pert1stO}. $\delta\pi_\phi^0(\phi^J)$ and $\delta V(\phi^J)$ are the only corrections functions with nonnull contributions.

To have a good reference frame, a nonzero energy density is ultimately required. This energy density can be as small as desired, hence any infinitesimal perturbation around the homogeneity restriction in Eq. \eqref{WaveFunctionHomogeneous} can lead to a good reference frame. Therefore, if we assume a perturbation around \eqref{WaveFunctionHomogeneous} to be valid, one can consider null its leading order in which such perturbations vanish exactly.

Taking into account all these simplifications, the fluctuations in the kinetic energy reduces to the two-point function which can be ultimately simplified to
\begin{equation}\begin{split}\label{2PointFunctionMatter}
		\frac{\langle\delta\rho_{\text{kin}}(\phi^0,k_i)\,\delta\rho_{\text{kin}}(\phi'^0,k'_i)\rangle}{\rho^2_{\text{kin}}(\phi^0)}=&\ 4\,\frac{\langle\delta V(\phi^0,k_i)\,\delta V(\phi'^0,k_i')\rangle}{V^2(\phi^0)}+4\,\frac{\langle\delta\pi^0_{\phi}(\phi^0,k_i)\,\delta\pi^0_{\phi}(\phi'^0,k'_i)\rangle}{\pi^0_{\phi}\,{}^2(\phi^0)}
		\\&-8\,\frac{\langle\delta\pi^0_{\phi}(\phi^0,k_i)\delta V(\phi'^0,k'_i)\rangle}{\pi^0_{\phi}(\phi^0)\,V(\phi^0)}\,.\end{split} \end{equation}
All the terms in the later expression involve observables which are independent of any of the rod fields $\phi^i$, $i=1,2,3$ (neither multiplicatively or in derivatives) and in the considered approximation, the mean field has no dependence on these fields either. For this choice of mean field, the first term has been already computed in the previous subsection, where it has been shown that a scale-invariant power spectrum with small amplitude is obtained. The other two terms written in this form behaves the same and consequently, Eq. \eqref{2PointFunctionMatter} give a scale-invariant power spectrum for both, matter density perturbations and volume density perturbations, with the small amplitude still scaling as $1/N$, as expected due to generic macroscopic observables defined for many-particles states.
However, scale invariance can be broken when different types of corrections are taken in consideration. On first place, it has been discussed in the previous subsection that if deviations from exact homogeneity are considered for the mean field, then non-scale-invariant terms should be taken under consideration. 
With respect to matter density perturbations, since at some point one should consider that rod fields acquire a nonzero background energy density, then their contribution to the expressions for perturbations should be also considered. This computation of this contribution can be found in \cite{CosmologicalPerturbations}.
Besides, if the gradient energy that has been previously neglected is taken into account, scale invariance is not expected in general to be preserve, even for the homogeneous mean field restriction.
Certainly, more work is needed to verify whether the assumptions discussed here are dynamically justified, even in presence of more realistic matter fields.

The results outline a concrete formalism for deriving a power spectrum of cosmological perturbations directly from a theory of quantum gravity, however to bring results closer to observational tests, more work is required. Being cosmological perturbations seeded by the quantum fluctuations produced as the natural behaviour of the quantum gravity condensate, one plausible situation is that perhaps, the observed initial power spectrum of quantum fluctuations is a kinematical property at a given time.
However, this primordial quantum fluctuations must be converted into classical perturbations during the propagation, “freeze out” and amplification when de universe expands.
Ref. \cite{InhomogeneousUniverse}, for instance, provides a physical mechanism for the emergence of a slightly inhomogeneous spacetime, computing the transition from the initial quantum fluctuations present in the deep quantum gravity regime to the usual gauge-invariant variables for observable scalar perturbations in cosmology.
This provides one more step towards a connection with the potentially observable spectrum of classical perturbations from full quantum gravity.

As mentioned before, the definition of volume perturbations depends in general on the chosen gauge, and its relation to the density perturbations and the curvature perturbation variable can be expressed through the gauge-invariant “curvature perturbation on uniform-density hypersurfaces” (see for instance \cite{Baumann})
\begin{equation}\label{CurvaturePerturbation}
	\zeta=\Phi+\frac{H}{\dot{\rho}}\delta\rho\end{equation}
In the later formula, $\Phi$ is the Bardeen variable parametrizing scalar perturbations, $H$ is the Hubble parameter and $\rho$ is the background matter density with $\delta\rho$ their corresponding density perturbations. 
Although, different gauge choices permits different relations between the terms in the r.h.s. of Eq. \eqref{CurvaturePerturbation}, this gauge freedom is lost due to the introduction of the relational coordinates. In the GFT context reference scalar matter fields define a harmonic gauge \cite{BHs}. The full computation of the observationally relevant dimensionless power spectrum of the gauge-invariant curvature perturbation variable $\zeta$  in the GFT condensate approach can be found in \cite{InhomogeneousUniverse} and its expression at leading order is given by
\begin{equation}
	\Delta_\zeta^2(k)\sim \frac{m_{\text Pl} \,V}{24 \pi \,j_0{}^{3/2} M^4} \,k^3\ .\end{equation}
where magnitudes such as $\sigma_j$ has been eliminated in favor of the variables normally used in cosmology: $j_0$, the dominant spin; the Plank mass $m_{\text Pl}$; $V\propto a^3$ is the total volume of the condensate which is related to the scale factor defined implicitly in the Hubble parameter in Eq. \eqref{CurvaturePerturbation}; and $M$, the mass scale set by the energy density in the scalar field.
An interesting feature of this result is that the spectral index derived from this expression is $n_s=4$, which is consistent with semiclassical cosmological calculations in the context of QFT in curved spacetime. 
Nonetheless, the original setup of this cosmological scenario is very simple and only uses massless scalar fields, hence one does not expect any matching with the actual CMB observations, but to illustrate what one would expect from the standard formalism within quantum field theory on curved spacetime.
It is clear that a generalization to more realistic models with more complicated matter dynamics beyond free massless scalar fields is required to bridge the gap between observations of the early universe and the condensate formalism.

Before concluding this section, let us mention another possible approach to GFT condensate cosmology from which standard general relativity can be approximated by means of effective corrections of quantum gravitational origin. A class of thermal condensates can be constructed by making use of generalised equilibrium Gibbs states based on the maximum entropy principle, together with certain quantum many-body techniques \cite{KotechaOriti2018,ChircoKotechaOriti2019}. A tentative picture arising from this formalism models the universe as a finite temperature condensate phase representing the effective macroscopic spacetime, together with a static thermal cloud naturally encoding the corresponding quantum geometric statistical fluctuations over it. This model exhibits all the expected features, i.e. coherence, entanglement and statistical fluctuations in a given set of observables \cite{AssanioussiKotecha-vacuaandcondensates-2020}. However, it displays two main differences with respect to the non-thermal GFT condensate cosmology discussed in this review. On the one hand, it is expected for the early time phase of the universe to be dominated by the thermal cloud; on the other hand, the model recovers the expected cosmological dynamics at late times when the thermal part is dominated by the condensate, but in this context the latter is generated dynamically \cite{AssanioussiKotecha-vacuaandcondensates-2020,AssanioussiKotecha-condensates-2020}. Let us remark that all results in GFT cosmology are reproduced when the fluctuations are completely turned off, namely the thermal states can be consistently reduced to the “zero temperature” coherent states that have been used to obtain the effective description for the flat FLRW spacetime in previous sections. However, these statistical coherent states may bring further progress to GFT condensate cosmology program by offering a tangible and controllable way of incorporating perturbations in relevant observables. Specifically, \cite{AssanioussiKotecha-condensates-2020} computes additional correction terms in the evolution equations finding a higher upper bound on the number of e-folds ${\cal N}$ even without including interactions. However, the increase in ${\cal N}$ is still not sufficient to match the physical observations estimated at ${\cal N}\sim60$. It is expected that the condensate would be affected by the presence of the thermal cloud, hence the relaxation in the static approximation of the latter implies modifications during early phase in the previously studied GFT cosmological models \cite{FriedmannEmergent,GFTInteracting}, by altering the inflation rate.

\section{Discussion and Closing Remarks}\label{Closing Discussion}

Relevant aspects of GFT cosmological implications have been discussed in this review for the purpose of enriching the debate on the possible scenarios that a suitable theory of gravity could open in cosmology. The consistency of the GFT condensate cosmology has grown in the last couple of years because it could rely not only on the convergent results with models derived from LQG. Further developments of the GFT approach could provide more hints regarding the effective cosmological dynamics. Indeed, the latter is a central open research field in GFT and focuses on the attempt at providing some constraints to terms appearing in the GFT action \eqref{Action}. In this respect, a relevant issue lies in determining which choices of ${\cal K}$ and ${\cal V}$ (see Section \ref{QGmatterRF}) are required to obtain a suitable quantum gravity theory that recovers general relativity in the classical limit. We want to focus on two different research lines in GFT approaches studying the effective macroscopic dynamics that builds up the cosmological model and that can be of interest for those working on others than GFT methods. 

1) The first one according to which within the GFT condensate cosmology is possible to realize an early era of geometric inflation. This is a period of accelerated expansion in absence of an inflaton field and its associated {\it ad hoc} potential. A detailed study of the condition for inflation can be found in \cite{UniverseAsQGC}; however, in \cite{Sakellariadou2017} it is argued that the number of e-folds computed for the free theory --${\cal V}=0$ in the action \eqref{Action}-- suggests that such a geometric inflation cannot last sufficiently long to accommodate observational data. This implies that GFT cosmology in absence of interactions between building blocks cannot replace the standard inflationary scenario. Studies explore the implications of including these interactions, which is indeed a more natural and consistent scenario, as the quanta of geometry should be somehow `glued' among each others instead of being in a sort of diluted gas regime of tetrahedra. Therefore, the results obtained in \cite{GFTInteracting} may be able to give an alternative prescription on how to build a GFT model with specific type of interactions, such that in the semiclassical limit the desired properties of our homogeneous and isotropic universe are obtained as an emergent 3–geometry. Interestingly, in the interacting case, one can find a range of the parameter space for which the inflationary era last for sufficiently long. However, to obtain a successful scenario one needs to verify that there is no intermediate stage of deceleration between the bounce and the end of inflation. According to \cite{NonIsotropic2} a real-valued condensate field has solutions avoiding the singularity and also growing exponentially after a bounce, if and only if the GFT energy is negative. A discussion on the possible values of the parameter space of the interacting potential can be found in \cite{Sakellariadou2017}, together with the stability properties of the evolving isotropic system, giving rise to effective continuous and homogeneous 3–geometries built from many smallest and almost at building blocks of quantum geometry.
				
2) The second research line is based on the successful use of relational observables to extract an effective dynamics in the cosmological sector of GFT\footnote{Anyway, in order to translate the theory into a set of equations for cosmological observables, the addition of a scalar field variable is crucial, since it allows us to define within the full theory a set of relational observables with a clear physical meaning. In GFT approaches we find models including a scalar field as matter content. The use of matter reference frames is not new; it dates back at least to DeWitt proposal \cite{DeWitt} where coordinates are proposed to be constructed with convenient matter scalar variables (in \cite{ObservablesInEffGrav} an extended discussion can be found). More contemporary advances have been obtained by using dust matter to account for this effect. First insights have been proposed by Brown and Kuchař \cite{DustFrame} and generalizations to LQG have been developed in Ref.  \cite{DustLQG}. Relevant advances have been done also by Gielen \cite{CosmologicalPerturbations,BHs}. With regard to models constructed from the theories discussed in this review, the employment of a massless scalar field as a relational clock defining relational dynamics also appears in canonical LQG \cite{Domagala.Giesel.Kaminski2010} and LQC \cite{Ashtekar.lowski.Singh2006,MasslessScalarsLQG}.}. Particularly interesting consequences have been derived using condensate states of a GFT with a relational Hamiltonian $\hat{\cal H}$ generating the evolution with respect to a massless scalar `clock' field $\phi^0$. This approach, initially proposed for a toy model in \cite{HilbertToyModel}, postulates a canonical reformulation of the classical GFT action making use of classical “sametime” Poisson brackets and providing a new path to define a quantum theory for GFT. It is based in a deparametrized setting in which some degrees of freedom (in the current discussion the singled out variable in the domain of the GFT field that corresponds to the scalar field degrees of freedom) serve as `coordinates' parametrizing the remaining ones. Concretely, one defines `equal relational time' commutation rules for the fundamental operators corresponding to the GFT field  $\varphi$ and its conjugate momentum. This procedure of choosing a suitable matter degree of freedom as a `clock' before quantisation is performed at the level of the collective, coarse-grained description of the microscopic GFT degrees of freedom, hence providing a “proto-geometric” notion of relational dynamics \cite{Wilson2019}. Nonetheless, this approach shows the limit of postulating a “tempus ante quantum” structure, since it precludes access to the quantum properties of the sub-system chosen as a clock \cite{MarchettiOriti2020}\footnote{It would be interesting to contrast the latter deparametrized framework for a single clock with
a covariant setting in which one can choose different clocks, following the ideas of \cite{MultimpleClocks}. In this respect, recent work has been done by \cite{GielenPolaczek}.}. 

In our view, the above-mentioned relational standpoint represents a fruitful field of investigation, suggesting important conceptual implications. Indeed, if physical systems evolve with respect to internal dynamical degrees of freedom of the theory, notions of physical clocks and rods should be searched in the well-behaved fundamental degrees of freedom. Recent insights in this direction deepen the relational strategy to solve the problem of time in emergent gravity. For instance, in \cite{MarchettiOriti2020} it is addressed the contribution of the quantum properties of the relational clock to the effective dynamics and in particular, \cite{GielenMenndezClock2020} analyses how the clock's properties define the evolution and determine the resolution of the initial singularity. 

Furthermore, an intriguing conceptual implication of this relational approach is that spacetime and, hence geometry, dissolves in pre-geometric “atoms” and the recovery of gravity is postulated as an emergent phenomenon deriving from their collective behaviour. Examples of these pre-geometric structures can also be the spin networks of LQG, the simplicial (piecewise-flat) geometries of lattice quantum gravity and, as it was extensively discussed, the quanta of GFT which can be understood as either spin networks and simplicial geometries of lattice quantum gravity, or as the causal sets in causal dynamical triangulation among other possibilities.

Nevertheless, to achieve the resulting collective structures one requires one more step, namely one needs a suitable averaging/coarse-graining procedure from where approximately continuous and regular structures emerge and allow to label the evolution of other degrees of freedom in the theory. Within the GFT approach, this is commonly known as the ‘proto-geometric’ phase of the theory. Thus a suitable clock should be searched in the ‘proto-geometric’ regime. According to our discussion, massless scalar fields are attached to each GFT entity, implying for each “single-quantum time” a pre-geometric interpretation. This means that the existence of a large number of quanta brings the complicated problem of reconstructing collective, coarse-grained synchronized states with an internal variable that could be used as a “relational clock” \cite{MarchettiOriti2020}. Contrary to the classical context where the troublesome diffeomorphism-invariant set up is circumvent by using specific solutions involving special isometries to which preferred temporal and/or spatial directions can be associated, in the quantum setting this solution is not allowed. Thus, one has to deal with the absence of preferred temporal and/or spatial directions. Future research in GFT will show whether this picture is consistent and will further explore its implications for Quantum Gravity theory and cosmology.

\section*{Acknowledgements}
We would like to thank Marco De Cesare, Claus Kiefer, Isha Kotecha and Daniele Oriti for fruitful discussions and suggestions. We also thank the referee for their remarks which led to an improvement of the manuscript. The research leading to this paper has received funding from the European Union’s Horizon 2020 research and innovation programme under grant agreement No 758145.

\end{document}